\documentclass[12pt,fleqn]{article}

\pdfoutput=1
\usepackage{setspace}
\usepackage{float}
\usepackage{fullpage}
\usepackage{float}
\usepackage{wrapfig}
\usepackage{amsthm,dcolumn,graphicx,multicol,fancyhdr}
\usepackage{latexsym,graphicx,ifthen}
\usepackage{rotating,amsfonts, color,amssymb,amsmath,amscd,array,enumerate,afterpage}
\usepackage{hyperref}
\usepackage{bm}
\usepackage{titling}
\usepackage{algorithm}
\usepackage{algpseudocode}
\usepackage{subcaption}
\usepackage[superscript,biblabel]{cite}

\newcolumntype{3}[1]{>{\centering\arraybackslash}p{#1}}

\newcommand{\bigzero}{\makebox(0,0){\textbf{\Large0}}}

\def\keywords{\vspace{.5em}
	{\textbf{Keywords}:\,\relax%
}}

\title{Assessing Health Care Interventions via an Interrupted Time Series Model: Study Power and Design Considerations }
\author{Maricela Cruz\thanks{Department of Statistics, University of California, Irvine, USA} \,, Daniel L. Gillen$^{*}$, Miriam Bender \thanks{Sue \& Bill Gross School of Nursing, University of California, Irvine, USA} \,, Hernando Ombao$^{*}$ \thanks{Statistics Program, King Abdullah University of Science and Technology (KAUST), Saudi Arabia} }

\date{\vspace{-3cm}}

\begin{document}
	
	\maketitle
	\doublespacing
	
	\vspace{.8cm}

\begin{abstract}
The delivery and assessment of quality health care is complex with many interacting and interdependent components. In terms of research design and statistical analysis, this complexity and interdependency makes it difficult to assess the true impact of interventions designed to improve patient health care outcomes. Interrupted time series (ITS) is a quasi-experimental design developed for inferring the effectiveness of a health policy intervention while accounting for temporal dependence within a single system or unit. Current standardized ITS methods do not simultaneously analyze data for several units, nor are there methods to test for the existence of a change point and to assess statistical power for study planning purposes in this context. To address this limitation, we propose the `Robust Multiple ITS' (R-MITS) model, appropriate for multi-unit ITS data, that allows for inference regarding the estimation of a global change point across units in the presence of a potentially lagged (or anticipatory) treatment effect. Under the R-MITS model, one can formally test for the existence of a change point and estimate the time delay between the formal intervention implementation and the over-all-unit intervention effect. We conducted empirical simulation studies to assess the type one error rate of the testing procedure, power for detecting specified change-point alternatives, and accuracy of the proposed estimating methodology. R-MITS is illustrated by analyzing patient satisfaction data from a hospital that implemented and evaluated a new care delivery model in multiple units.

\end{abstract}


\thispagestyle{empty}

\setcounter{page}{1}
\section{Introduction}

The delivery and assessment of quality health care is increasing in complexity. Now more than ever, patients, providers, resources and contexts of care interact in dynamic ways to produce various measurable health outcomes, that oftentimes do not align with expectations. \cite{Hawe:2015ca} 
Assessing the impact of health interventions on patient health outcomes 
is therefore inherently difficult with regards to research design and statistical analysis. \cite{Datta:2013gf, campbell2000}  
Interrupted time series (ITS) designs borrow from traditional case-crossover designs and function as quasi-experimental methodology that allows each sampling unit to serve as its own control without stripping contextual and temporal factors from the analysis. \cite{Shadish:2002uv}
Current standardized methods for analyzing ITS designs do not borrow information across units. 
This is a serious limitation because it does not take advantage of all available data that may provide information on the lag associated with a given intervention. A main contribution of the work presented here are the empirical power studies that illustrate the gain in efficiency obtained by borrowing information across units. 

\begin{figure}[!htbp] 
 	\centering
	\includegraphics[scale=.18]{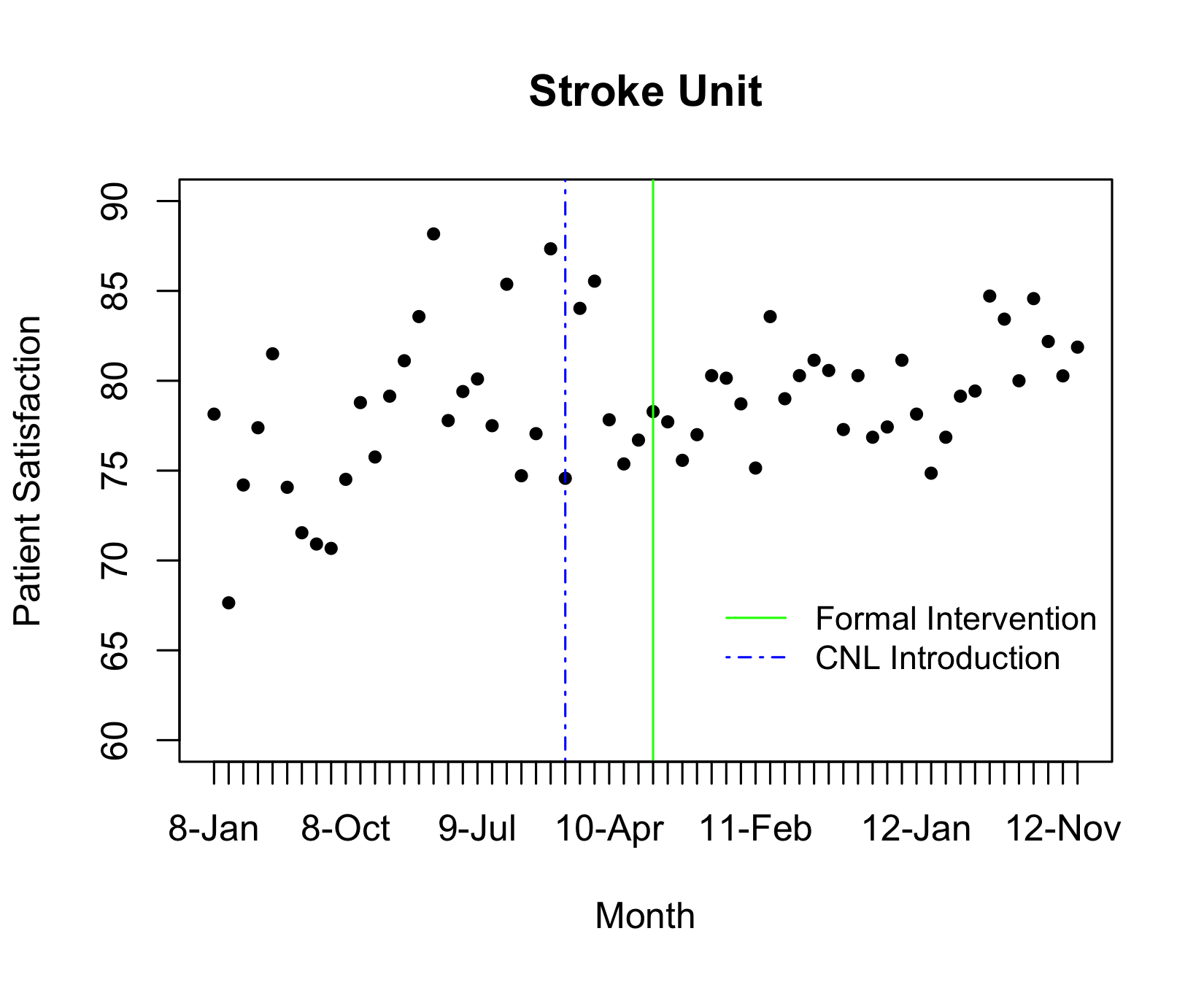}
	\includegraphics[scale=.18]{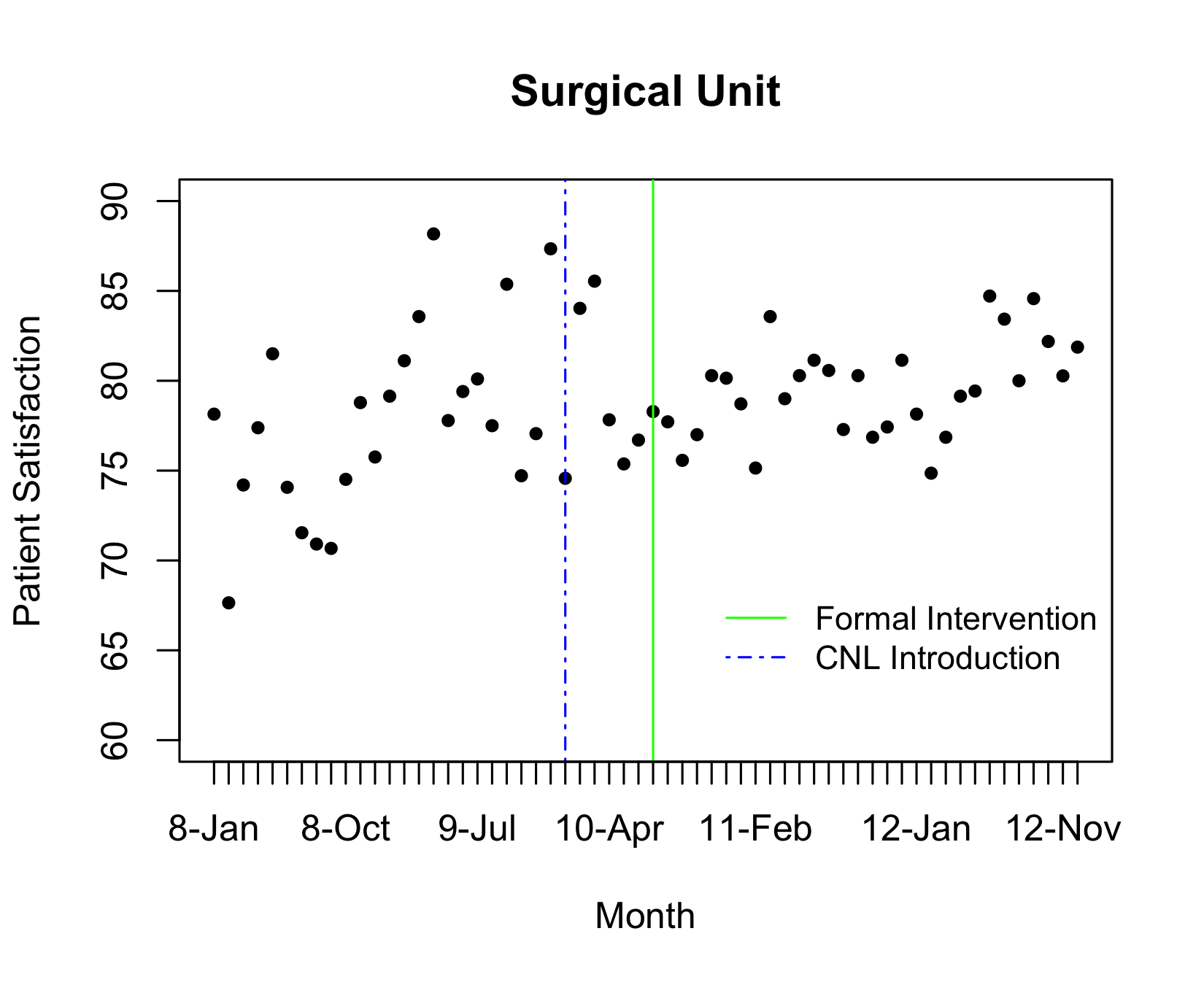} 
	\vspace{-.2cm}
	\caption{Plots the time series of observed average patient satisfaction for the Stroke and Surgical units.}
		\label{UnitTS}
\end{figure}

The methodology presented in this paper is motivated by our interest in estimating the lagged effect of an intervention on average patient satisfaction survey scores, recorded monthly at five clinical care units. A time series plot of patient satisfaction scores from January 2008 to December 2012 at two hospital units (the Stroke and Surgical units), is given in Figure \ref{UnitTS}.  There seems to be a change in the mean functions of the Stroke and Surgical units around the middle of the time series, slightly before the formal implementation of the intervention on July 2010.
The time series data are from a study aimed to assess the impact of a new nursing care delivery system on publicly recorded standardized quality and safety metrics. \cite{ClinicalNurseLeade:2015tr} 
These metrics are a central area for improvement because the Center for Medicaid and Medicare Services (CMS) Value Based Purchasing Program utilizes them for health systems' care services reimbursement. \cite{Kavaanagh:2012dt}

The intervention was the implementation of Clinical Nurse Leader (CNL) integrated care delivery, a nursing model that embeds a master prepared nurse into the front lines of care. \cite{Bender:2017} The nurses, referred to as Clinical Nurse Leaders (CNLs), have advanced competencies in clinical leadership, care environment management, and clinical outcomes management. \cite{Bender:2017} The CNLs, conducting their master's level microsystem change project, were introduced into their respective hospital units on January 2010, six months prior to the formal intervention implementation time.
This may or may not have influenced the ‘change point’ of the intervention effect. Namely because of this early introduction, the estimated change point may have occurred up to 6 months prior to the formal intervention time point. We are therefore interested in estimating the time lag (or delay) between the onset of the intervention --- i.e., the change point --- and the effect on patient satisfaction. 
Our model assumes a global change point rather than unit-specific change points (1.) to pool information across hospital units and increase efficiency, and (2.) to reduce the impact of unit specific high-leverage points around the CNL and formal intervention implementation time points. 
Importantly, we are interested in examining whether or not a change point actually exists, thereby deducing whether or not the intervention impacts patient satisfaction. Our interest is not solely on properly modeling the CNL intervention; we are also interested in future study designs, and so, focus on power.

The most utilized statistical methodology for analyzing interrupted time series data in the health care literature is segmented regression. \cite{Taljaard:2014kc, Penfold:2013bc, wagner2002, linden2015} Segmented regression restricts the analysis to one health care outcome for one unit (group or cluster). 
In the context of assessing the above intervention, perhaps a severe drawback of segmented regression is that it restricts the interruption to a predetermined time point in the series or censors data by removing the set of time points for which the intervention effects may not be realized. Additionally, segmented regression neglects the plausible differences in autocorrelation and variability between the pre- and post-intervention phases present in the data.
The Robust-ITS model proposed by Cruz, Bender, and Ombao (2017)
treats the change point as variable, appropriate for situations where the data warrants such treatment, and tests for differences in autocorrelation 
and variability pre- and post-change point. \cite{Cruz2017} Nevertheless, 
Robust-ITS and segmented regression both neglect shared information across hospital units and inherently assume a change point \textit{exists}. 

Assessing the impact of an intervention with traditional segmented regression or Robust-ITS on these data requires a separate analysis for each individual unit 
We expect many of the units to share several characteristics --- i.e., abide by the same regulations, have similar schedules, hire staff based on the same criteria, etc. --- because the units are housed within one hospital. Moreover, we expect the CNL `training' or education for each of the CNL students to include commonalities, such as course work and care delivery ideology. Assessing the intervention impact on multiple units via current segmented regression methods ignores shared characteristics across units, in particular the similarity between characteristics influencing the change-point.

Inherently assuming that a change point exists, as in segmented regression and Robust-ITS, may lead to erroneous results when there is no actual a change point.
Change point models will forcefully quantify a change in the outcome regardless of the presence of a true change point. This is a problem whether the change point is determined a priori or estimated over a set of possible change points. Assuming a change point exists when it truly does not, will force a model to provide an estimate of an artificial difference in the outcome. To avoid incorrectly specifying an unnecessary change point and regression to the mean phenomena, we focus on formally testing for the existence of a change point.

In this paper, we develop the Robust Multiple ITS model (R-MITS), a novel extension of Robust-ITS, appropriate for multiple independent interrupted time series. Furthermore, we present the supremum Wald test, able to test for the existence of a change point across units. Importantly, we provide empirical type one error, power, and accuracy studies assessing the operating characteristics of our developed methodology. 
The proposed method (a.) borrows information across hospital units to increase efficiency, (b.) estimates a global change point of an instituted intervention, (c.) formally tests for the existence of a change point in the unit specific mean functions, and (d.) allows for changes in the mean functions and autocorrelation structures across units. 

We go on to describe our proposed R-MITS model and provide details on the estimation and inference procedures. In our model specification we outline the supremum Wald test used to determine the necessity of a change point. Next, we present empirical simulations to assess the type one error, power for detecting specified change point alternatives, and accuracy of the change point estimation procedure. We then analyze the impact of Clinical Nurse Leader integrated care delivery on patient satisfaction. Lastly, we present a summary of our developed methodology and briefly describe future work. 

\section{The Robust Multiple ITS Model} 

Our proposed model tests for the existence --- rather than merely assume --- of a change point and 
adequately manages multiple units/time series.
A noteworthy feature of our approach is the clear distinction between the time of intervention and the change point, as in Robust-ITS. 
Setting the change point to a predetermined time may lead to incorrect measures of the intervention's effect on the system; particularly when set to the intervention time, because that does not necessarily represent the reality that complex interventions may have varied effects and take time to manifest change.
Prevalent approaches to overcoming this limitation are to remove, or censor, a specific set of time points from the analysis. \cite{Penfold:2013bc, Taljaard:2014kc} 
R-MITS borrows information from all microsystems to estimate a global change point; i.e., determines the time point at which the effect of the intervention initiates for the entire health system.  
Moreover, detecting differences in autocorrelation and variances pre- and post-intervention is critical in evaluating the effectiveness of an intervention.
The R-MITS model allows for two completely different data dependency and variability structures to exist prior to the intervention and post-intervention within each unit. 

To prelude model development we plot the outcome against time to (a.) illuminate the functional form of the longitudinal mean over time; (b.) determine the presence of seasonality, and (c.) further investigate the set of plausible change points and the necessity of a change point. 
If the functional form of the longitudinal mean is not linear, we transform the outcome to obtain a linear pattern, or apply a different segmented regression model appropriate for the pattern present within the ITS design. When needed, we account for seasonality via traditional statistical methods concisely described in Bhaskaran et al. (2013). \cite{bhaskaran2013} Although not used in the analyses here, variance stabilizing transforms can be applied on the outcomes of interest if necessary.
In our interrupted time series data, the longitudinal mean functions are relatively linear in time with no apparent seasonality. Thus, no transformations are applied on the outcomes of interest.

\subsection{Description of R-MITS} \label{Section3}
 
Denote $t^*$ as the time point at which the intervention is introduced and $\tau$ as the time point at which the effect of the intervention initiates (the change point) for the outcome of interest. Sometimes it may indeed be true that $t^* = \tau,$ but this may not necessarily be true for all outcomes. Often it is entirely possible that
the time of effect of the intervention differs from the time of intervention introduction (i.e., either $\tau > t^*$ or $\tau < t^*$). If  $\tau > t^*$  then the effect of the intervention on the outcome is not realized until after the formal intervention time point. As might be the case when a learning effect exists with regards to the intervention, thereby leading to a delay in the realization of the full intervention impact.   
When $\tau < t^*$ there is an anticipatory intervention effect on the outcome. This may be the case in our motivating study, where Clinical Nurse Leaders are introduced into units prior to the formal intervention start time. 
We propose a data adaptive procedure for estimating and determining the existence of $\tau,$ discussed in Section \ref{CPSection}. 
Many change point detection methods in time series exist, but often deal only with changes in the mean functions and variance (not the autocorrelation structure itself), and may not work well in shorter time series. \cite{davis2006, Ombao:2015} The method proposed in this paper can suitably manage changes in the autocorrelation structure, as well as in the mean functions and volatility.

Define $y_{jt}$ as the outcome of interest for hospital unit $j$ at time $t$ (where $j =1, \dots, J$ and $t=1, \dots, T$). For 
example, $y_{jt}$ may be patient satisfaction for the Stroke unit at time $t$. 
The general regression is defined as
\begin{equation} \label{main}
 y_{jt} = \mu_{jt} + \epsilon_{jt}, \end{equation} where
$\mu_{jt}$ is the mean function and $\epsilon_{jt}$ is the 
stochastic process that models that fluctuations around the mean function.
The mean component, $\mu_{jt}$, characterizes the mean function of the response for unit $j$ during the pre-intervention and post-intervention phases.
The stochastic process, $\epsilon_{jt}$, accounts for the variability and correlation of the outcome in the $j^{\text{th}}$ unit. 
In the following discussion we define the mean functions and stochastic components for the R-MITS model, and the estimation procedures.

\subsubsection{The Pre- and Post-Intervention Mean Function}\label{Mean}

The mean function of the outcome for hospital unit $j$ at time $t$ is
\begin{equation} \label{EQ3}
\mu_{jt} = \left\{
\begin{array}{lr}
\beta_{j0} + \beta_{j1} \, t,  &  \,  t <  \tau \\
( \beta_{j0} + \delta_{j}) \, + (\beta_{j1} + \Delta_{j}) t,  &  t \geq \tau 
\end{array}
\right. .
\end{equation}
The parameters in $\mu_{jt}$ are: (1.)  $\beta_{j0},$ the intercept of the mean function prior to the change point; (2.) $\beta_{j1},$ the slope of the outcome prior to the change point; (3.) $\beta_{j0} + \delta_j,$ the intercept of the post-intervention phase; (4.) $\beta_{j1} + \Delta_j,$ the slope of the post-intervention phase; for the outcome in unit $j$, and (5.) $\tau,$ the global over-all-unit change point of the response. Thus, $\delta_{j}=\Delta_j=0$ implies there is no change in the mean structure before and after time $\tau$.  Health care specialists are primarily interested in testing for the intervention lag (delay in the effect of the intervention), and the differences in the outcome means between the pre- and post-change point phases.

\vspace{0.07in}
\noindent {\bf Remark (1.)} The metrics adopted by the health policy evaluation literature to assess the effect size of an intervention via ITS designs are the change in level and change in trend (or slopes). While the level change identifies the size of an intervention's effect, the change in trend quantifies the impact of the intervention on the overall mean function. It is necessary to report both level change and change in trend to interpret the results of an ITS study accurately. \cite{EPOC:2015ww} 

\begin{figure}[!htbp] 
	\centering
	\includegraphics[scale=.22]{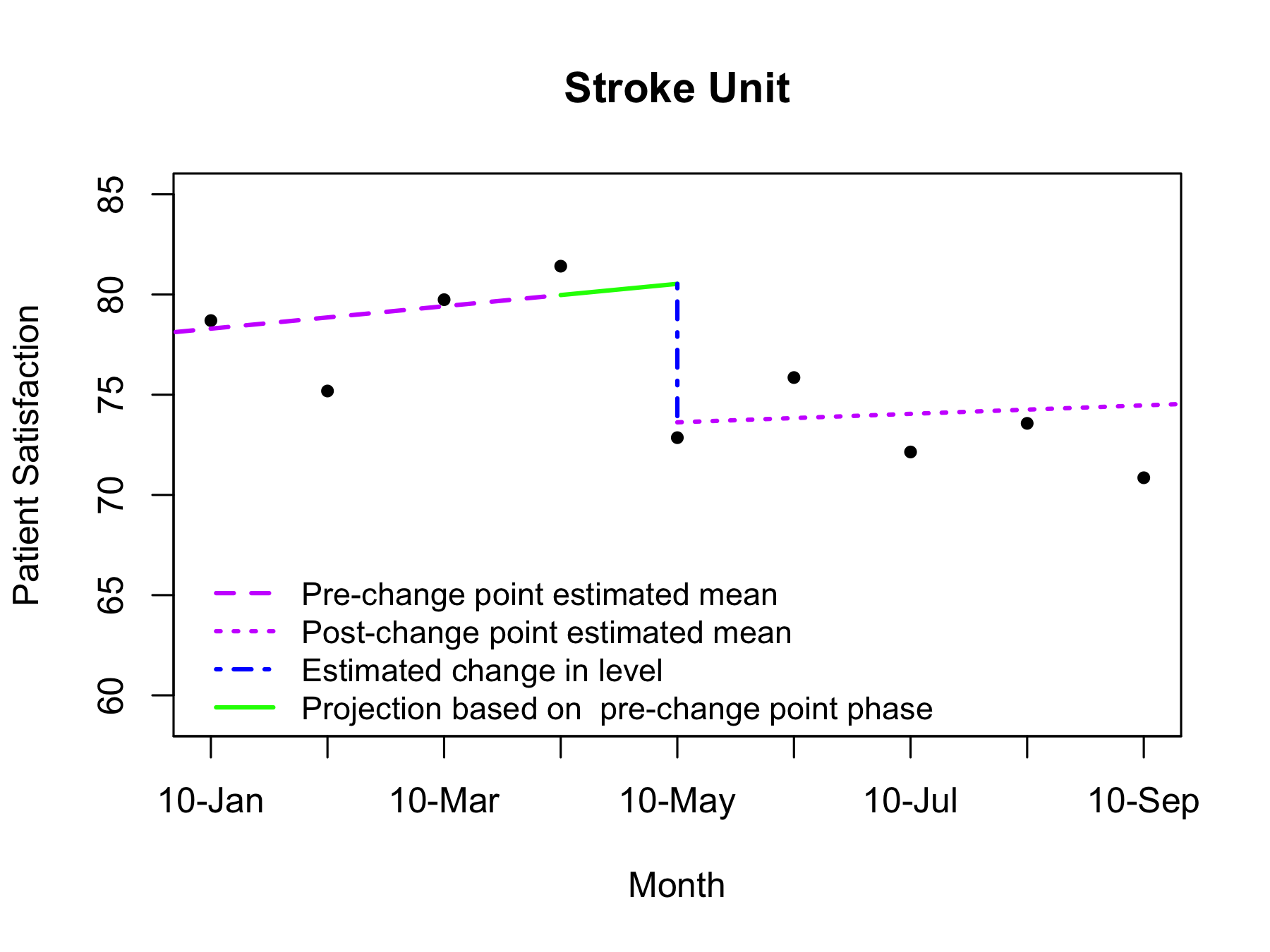} 
	\caption{An example of a segmented regression model fit for the Stroke unit. The plot depicts (1) the segmented regression lines fit to the pre- and post-change point phases, (2) the projection of the mean at the change point based on the pre-change point regression, and (3) the change in level as defined here. The plot contains data from January 2010 to September 2010, instead of the entire observational period, to clearly illustrate the level change.}
	\label{CIL}
\end{figure}

\vspace{0.07in}
\noindent {\bf Remark (2.)}  The level change is interpreted as the change in the anchored intercept (anchored at the change point), and is therefore the jump between the projected mean function based on the pre-change point phase and the estimated mean function post-change point. In our model the unit specific change in level is defined mathematically as $\delta_j + \Delta_j \tau,$ and is graphically depicted in Figure \ref{CIL}. Trend change, or slope change, is denoted by $\Delta_j$ in the mean function, equation (\ref{EQ3}).

\vspace{0.1in}

The mean function parameters are estimated simultaneously with the stochastic component parameters and change point, via maximizing the conditional likelihood given in equation (\ref{CondLike}) of Section \ref{CPSection}, with the auto-regressive coefficients' estimator accounting for the volatility of the shifted series.  
with variance components  shifted AR(1)  An algorithm on how to precisely estimate the parameters is provided in Section \ref{CPSection}.

\subsubsection{Stochastic Properties Pre- and Post-Intervention}

The stochastic component in equation (\ref{main}), $\epsilon_{jt}$, captures the autocorrelation structure of the outcome variable across time for unit $j$, and may change as a result of the intervention; the $\epsilon_{jt}$ are zero-mean random fluctuations around the mean function of unit $j.$
One goal of the CNL intervention is to increase the consistency of care delivery and hence patient assessment outcomes, (i.e., decreasing variability of the outcomes). 
We therefore include separate stochastic components for the pre- and post-change point phases, to allow for a change in outcome variability.

Due to the impact of the intervention, the stochastic process pre-intervention might differ from the stochastic process post-intervention. 
That is, $\epsilon_{jt}$ for $t \in \{1, \dots, \tau -1 \}$ may be a different stochastic process than $\epsilon_{jt}$ for $t \in \{ \tau, \dots, T\}.$ Note, the length of the time series is denoted by $T.$ Hence, the autocorrelation and variance might differ pre- and post-change point. Here, the stationarity requirement is satisfied if the variance, mean function, and autocorrelation are constant within each stochastic process, not constant across all time points.

In order to fit stationary AR or ARMA processes to the stochastic components, one should first confirm that there are no striking signs of non-stationarity. That is, the mean and variance of the residuals (obtained from modeling and removing the mean function as in the previous section) must be relatively constant. If the mean function is not misspecified, then the residuals should be fluctuating around zero without trend. Moreover, the residuals should be stationary within each of the pre- and post-intervention phases. \cite{Shumway:2011}
Our analysis of patient satisfaction suggests that it is reasonable to assume stationarity within each phase, and hence we proceed with the assumption of stationarity. 

In this work we use the AR(1) process to model the stochastic component, $\vec{\epsilon}_{j} = Y_{j} - \vec{\mu}_{j}$, where $Y_{j} = [y_{j2}, \dots,  y_{jT} ]^{'}$ and $\vec{\mu}_{j} = [\mu_{j2}, \dots,  \mu_{jT} ]^{'}$ for unit $j.$ Note, $y_{j1}$ is not included in $Y_{j}$ and $\mu_{j1}$ is not included in $\vec{\mu}_j$, because we condition on the first observation. The AR(1) coefficient is estimated by maximizing the conditional likelihood with the denominator of the estimator averaging the volatility of the shifted AR(1) series. We therefore condition on the first observation $y_{j1}.$
Since the mean function $\vec{\mu}_{j}$ is not known (we only have its estimate, $\widehat{\vec{\mu}}_{j}$), the stochastic component is not directly observed. Hence, we use the residuals $R_{j} = Y_{j} -  \widehat{\vec{\mu}}_{j} \equiv [r_{j2}, \dots, r_{jT}]^{'}$  \, in place of $\vec{\epsilon}_{j}.$ 
The residuals are modeled as:
\begin{equation} \label{AR1}
r_{jt} = \left\{
\begin{array}{lrr}
\phi_{j1} \, r_{j, t-1} + e_{jt, 1},  &  & \,\, \, \,  \, \, \,\, \, \,  \, \, \,  \, \, \, \, \, \, \, \, \,  \, \, \, 1 \, < \, t \, \leq \, \widehat{\tau} -1\\
\phi_{j2}  \, r_{j, t-1} + e_{jt, 2}  &   &  \widehat{\tau} -1 \,  < \, t \, \leq \, T. 
\end{array}
\right.
\end{equation}
To ensure causality in the time series sense, $\phi_{j1}$ and $\phi_{j2}$ must lie in the interval $(-1, 1)$ for all $j.$
Note, the auto-regressive coefficient prior to the change point, $\phi_{j1}$, is the correlation between time point $t$ and $t+1$ (the adjacent correlation or autocorrelation) where $t$ and $t+1$ belong to the pre-change point phase ($t, \text{and  } t+1 \in \{ 1, \dots, \tau -1 \}$), and $\phi_{j1}^{\vert h \vert }$ is the correlation between two time points $h$ units away (say $t$ and $t+h,$ both in the pre-change point phase) of the outcome. The auto-regressive coefficient post-change point, $\phi_{j2}$ has a similar interpretation. The zero-mean random fluctuations of model (\ref{AR1}) are white noise, $e_{jt, i} \stackrel{iid}{\sim} N \big( 0, \sigma_{jw, i}^2 \big)$ for $i \in \{1, 2\}.$ The variance of the distribution of the response at any time point $t$ is $ \sigma_{ji}^2 = \frac{ \sigma_{jw, i}^2}{1 - \phi_{ji}^2}$ for $i \in \{1, 2\}.$

The variance and auto-regressive coefficients in the AR(1) setting can be estimated by maximizing the conditional likelihood provided in equation (\ref{CondLike}) of Section \ref{CPSection}, with the auto-regressive coefficients' estimator accounting for the volatility of the shifted series.  
The structure of the variance-covariance matrix, and the estimators of the auto-regressive coefficients and white-noise standard deviations are given in the Appendix. 

To determine whether the stochastic process differs as a result of the change point for each unit, one can test the hypothesis that $\nu_j \equiv \phi_{j2} - \phi_{j1}$ equals zero. This can be tested by either estimating $\nu_j$ directly or by conducting an F-test for nested models. The F-test for nested models for the AR(1) scenario is described by Cruz, Bender, and Ombao (2017). \cite{Cruz2017}

\subsubsection{Estimation of the Change-Point and Model Parameters}\label{CPSection} 

In this paper we propose a conditional likelihood procedure for estimating the global change point. The set of possible change points is established by the researcher. We estimate the change point and therefore all of the parameters, both from the mean functions and stochastic components, simultaneously by obtaining the generalized least squares estimates.
Then we test for the existence of a change in the mean functions --- i.e., we test the null hypothesis that there is no change in any of the mean functions versus the alternative that there is a change in at least one of the mean functions --- at each possible change point by applying the supremum Wald test, described in section \ref{Wald}.

Define the length of the time series as $T$, the number of units as $J,$ the vector of mean function parameters as $\theta_j =  [ \beta_{j0}, \, \beta_{j1}, \, \delta_j, \, \Delta_j ]'$ and ${\bm \Sigma}_{j}$ as the variance-covariance matrix of the response in unit $j.$ The structure of the variance-covariance matrices is included in the Appendix.  

Let $q$ be a candidate change point in the set of possible change points $Q,$ where $Q = \{ t^* - m, \dots, t^*, \dots t^* + k \}$ for positive integer values of $m$ and $k$ set by the researcher. Recall the response vector for unit $j$ is $Y_{j} = [y_{j2}, \dots,  y_{jT} ]^{'}.$ Note, $y_{j1}$ is not included in $Y_{j}$ because we model the zero-mean random fluctuations around the mean functions as AR(1) processes. 
For each candidate change point $q \in Q$ we derive the conditional likelihood function, conditional on the first observations,
\begin{equation} \label{CondLike}
\begin{split}
L \, & \big(\theta_1, \, {\bm \Sigma}_{1} \,, \,  \dots,  \, \, \theta_J, \, {\bm \Sigma}_{J} \, \,  \vert \, q , \,  Y_{1}, \, \dots, Y_{J} \, \big) \\
& \equiv  \prod_{j=1}^{J}  \, \bigg( \frac{1}{\sqrt{2 \pi}} \bigg)^{T-1} \,\, \vert {\bm \Sigma}_{j} \vert^{-\frac{1}{2}}  \,\,  \exp{ \bigg\{ - \frac{1}{2} \big( \, Y_{j} - {\bm X(q)}_j  \,\,  \theta_j \, \big)^{'} \, \big({\bm \Sigma}_{j} \big )^{-1} \big( \, Y_{j} - {\bm X(q)}_j \, \, \theta_j \, \big) \bigg\} },
\end{split}
\end{equation}
where
$$
 \underset{(T-1)\times4}{{\bm X(q)}_j }\,  \, \,\equiv \begin{bmatrix}
1 & 2 &  0 & 0 \\
\vdots & \vdots & \vdots & \vdots \\
1 & q-1 & 0 & 0 \\
1 & q & 1 & q \\
\vdots & \vdots & \vdots & \vdots \\
1 & T & 1  & T 
\end{bmatrix}.
$$
We iteratively estimate $\underset{4\times1}{\theta_j}$  and $\underset{(T-1) \times (T-1)}{{ \bm \Sigma}_{j}}$ for all $j$, as in Algorithm 1. 

\begin{algorithm}[H]
	\caption{Estimating $\theta_j$ and $\bm{\Sigma}_{j}$ iteratively}
	\begin{algorithmic}[1]
		\For{ $j \in \{1, \dots, J\}$}
		\State set $\zeta = 1$
		\State set $\widehat{\theta}_j^{0}$ to OLS estimates 
		\State from residuals $R_{j}^{0}$ calculate \, $\widehat{\phi}_{j1}^{0},$ \, $\widehat{\phi}_{j2}^{0},$  \, $\widehat{\sigma}_{j1}^{0},$ \, and \, $\widehat{\sigma}_{jw, 2}^{0}$ \, and generate $\widehat{{\bm \Sigma}}_{j}^{0}$
		\While{ $\zeta > $  tol \, }
		\State set $i$ to the iteration
		\State calculate $\widehat{\theta}_j^{i}$ based on $\widehat{ {\bm \Sigma}}_{j}^{i-1}$
		\State use residuals  $R_{j}^{i}$  to estimate  $\widehat{\phi}_{j1}^{i},$ \, $\widehat{\phi}_{j2}^{i},$  \, $\widehat{\sigma}_{jw, 1}^{i},$ \, and \, $\widehat{\sigma}_{jw, 2}^{i}$ \,
		\State obtain $\widehat{\bm{\Sigma}}_{j}^{i}$,
		\State set $\zeta$ to the Euclidean distance between $[ \widehat{\phi}_{j1}^{i -1}, \, \widehat{\phi}_{j2}^{i-1}]$ and $[\widehat{\phi}_{j1}^{i}, \, \widehat{\phi}_{j2}^{i}]$
		\EndWhile
		\EndFor
	\end{algorithmic}
\end{algorithm}

Define 
$$ L(q) \, =  \underset{(\theta_1, \, {\bm \Sigma}_{1} \,, \,  \dots, \, \, \theta_J, \, {\bm \Sigma}_{J})}{\max} \,\,\,\, L \, \big(\theta_1, \, {\bm \Sigma}_{1} \,, \,  \dots, \, \, \theta_J, \, {\bm \Sigma}_{J} \, \,  \vert \, q , \,  Y_{1}, \, \dots, Y_{J} \, \big),$$ 
then the estimated change point is
$$\widehat{\tau} =  \arg \max_{ q \in Q} \, L(q).$$
The estimates of $\theta_1, \dots, \theta_J$ are the generalized least squares (GLS) estimates obtained, after the desired tolerance level is reached, conditional on $\widehat{\tau}.$
The GLS estimates of $\theta_j$ and ${\bm \Sigma}_{j}$ for all $j$ given $\widehat{\tau}$ are
$$ [ \,\{ \widehat{\theta}_1,\, \, \widehat{{\bm \Sigma}}_{1} \}, \, \,  \dots,  \, \, \{\widehat{\theta}_J,\, \, \widehat{{\bm \Sigma}}_{J} \, \} \,] \, \, \, =  \underset{ [ \{\theta_1,\, \, {\bm \Sigma}_{1} \}, \, \,  \dots \, ,  \, \{\theta_J,\, \, {\bm \Sigma}_{J} \} ] }{\arg \, \, \max} \, L \big( \theta_1, \, {\bm \Sigma}_{1}, \, \dots, \theta_J, \, {\bm \Sigma}_{J}  \, \vert \, \widehat{ \tau }, Y_{1}, \, \dots, Y_{J}\big).$$ 

The presence of $\tau$ does not restrict the model to a fixed interruption with an instantaneous effect. In fact, $\tau$ allows the design matrix and estimates to transform based on the information the data provides. Importantly, the inclusion of an over-all-unit change point allows us to utilize information from all available units to determine when the intervention begins to affect the outcome globally. 
This flexibility of the model can be helpful in minimizing misleading results from an assumed change point.

\subsubsection{Multivariate Wald Test for the Existence of a Change Point} \label{Wald}

The change point is estimated by maximizing the conditional likelihood over the set $Q,$ and thus, concurrently estimates all other model parameters at each possible change point. Since we test for the existence of a change point at each $q \in Q,$ multiple testing bias exists if one utilizes standard critical values.  As such it is necessary to apply a correction to control the family-wise type I error rate. To this end, we calculate the multivariate Wald test statistic for every $q \in Q.$ We apply the Benjamini-Hochberg method to adjust for the total number of tests conducted --- the total number of tests is equal to the cardinality of $Q.$ The Benjamini-Hochberg method controls the false discovery rate; control of the false discovery rate weakly implies control of the family wise type 1 error rate for an $\alpha =0.05$ level. \cite{Benjamini}
In this case, a binary decision of whether a change point exists or not corresponds to a rejection of the null hypothesis for any one of the tests conducted. 

We focus on determining the existence of a change point across the unit specific mean functions, i.e., for each $q \in Q$ we test 
\begin{align*}
H_0 &: \,  \delta_j=\Delta_j=0 ~~ \forall ~j, ~j=1,\dots,J\text{ (no change point)} \\
\text{vs.} ~~H_a &:  \,  \,  \delta_j \ne 0 \text{ and/or } \Delta_j \ne 0, ~ \text{for some} ~j,  ~j=1,\dots,J,  ~  \text{ (a change point at} ~q  \text{)} .
\end{align*}

Even though our model assumes a global change point to pool information across units for efficiency, a rejection of the null hypothesis for our Wald test implies a change point in at least one of the hospital units. A rejection does not imply that a change point exists across all units and is the same in all hospital units. Moreover, we do not restrict the impact of the change point at each unit --- i.e., we allow the change in level and change in slope to differ across units as in R-MITS. We borrow information across units for the estimation of the global change point, but we do not force the impact on the outcome to be the same in each unit.
Clearly, if one wanted to establish the existence of a change point for a particular unit, enough data would have to be gathered within that single unit to detect and estimate (with high enough precision) the unit specific change point.

The hypotheses can be written in terms of full and reduced mean function models.
Define the full and reduced mean function models as
\begin{gather} 
\mu_{jt}^{1} = \beta_{j, 0} + \beta_{j, 1} \, t \, + \, \big(  \delta_j + \Delta_j \, t \big) I( t \geq q), \label{FullModels} \\
\mu_{jt}^{0} =  \beta_{j, 0} + \beta_{j, 1}\, t, \label{NullModels}
\end{gather}
respectively. The full model, appropriate under the alternative hypothesis, is essentially the model of equation (\ref{EQ3}) and models a change in the mean functions at $q.$ The reduced model, appropriate under the null hypothesis, assumes the same mean functions across the entire observational period. 
Based on our model specification, testing the above hypotheses is equivalent to testing 
\setlength{\arraycolsep}{2.5pt}
\medmuskip = 1mu
\begin{gather*}
H_0 : \,  \begin{bmatrix}
\delta_1 \\ \Delta_1 \\ \vdots \\ \delta_J \\ \Delta_J 
\end{bmatrix} =  \begin{bmatrix}
0 \\ 0 \\ \vdots \\ 0 \\ 0
\end{bmatrix} \,\,\, \text{vs.} \,\,\,\,\,
H_a :  \,  \ \begin{bmatrix}
\delta_1 \\ \Delta_1 \\ \vdots \\ \delta_J \\ \Delta_J 
\end{bmatrix} \neq  \begin{bmatrix}
0 \\ 0 \\ \vdots \\ 0 \\ 0
\end{bmatrix}.  \,\,\,\,\, 
\text{Let} \,\, \mathbf{C} = \left[ \begin{array}{c@{}cc}
\begin{array}{cccc}
0 & 0 & 1 & 0 \\
0 & 0& 0& 1 \\
\end{array} & & \bigzero  \\
& \ddots & \\
\bigzero & & \begin{array}{cccc}
0 & 0 & 1 & 0 \\
0 & 0& 0& 1 \\
\end{array}
\end{array} 	\right],
\end{gather*}
$\vec{\beta}^{1} = [\beta_{1, 0}^{1} \,\,\,  \beta_{1, 1}^{1} \,\, \delta_1 \,\, \Delta_1 \, \dots \, \beta_{J, 0}^{1} \,\,\,  \beta_{J, 1}^{1} \,\, \delta_J \,\, \Delta_J ]^{'}$ and $\vec{\beta}^{ 0} = [\beta_{1, 0}^{0} \,\,\,  \beta_{1, 1}^{0} \dots \, \beta_{J, 0}^{0} \,\,\,  \beta_{J, 1}^{0} ]^{'}.$ 
Then the hypotheses can be written as 
$$
H_0 : \,  {\bf C} \, \vec{\beta}^{1} = \vec{0} \,\, \, \text{vs.} \,\, \,
H_a :  \, {\bf C} \, \vec{\beta}^{1}  \neq \vec{0},
$$ The multivariate Wald test statistic is given by
\begin{equation}
W = \big( {\bf C} \widehat{\vec{\beta}}^{1} \, \big) \, \big[{\bf C}  \,\, {\bf \widehat{V}} (\widehat{\vec{\beta}}^{0} ) \, {\bf C}^{'}\big]^{-1}  \big( {\bf C} \widehat{\vec{\beta}}^{1} \, \big)^{'} \, \overset{H_0}{\overset{.}{\sim}} \, \chi^2_{2J}, 
\end{equation}
where $ { \bf \widehat{V}}(\widehat{\vec{\beta}}^{0})$ is the block diagonal estimator of the variance covariance matrix of $\widehat{\vec{\beta}}^{0}.$ 
We specify  $ { \bf \widehat{V}}(\widehat{\vec{\beta}}^{0})$ in the Appendix. Note, we allow $\delta_j$ and  $\Delta_j$ to differ for each $j,$ i.e., for each unit.

We calculate the multivariate Wald statistic for each $q \in Q.$ Then we apply the Benjamini-Hochberg procedure to obtain corrected critical values. The Bejamini-Hochberg procedure is fully described by Benjamini and Hochberg (1995). \cite{Benjamini} 
If any of the multivariate Wald tests provide significant results, when compared to the corrected critical values, we conclude that a change point exists for at least one of the units.
The resulting `supremum Wald test' (SWT) is appropriate for detecting a change in any of the mean functions over a set of possible change points. Our test accounts for the heterogeneity of the mean functions and autocorrelation structures across units. In the following sections we illustrate that the SWT has empirically high power under specified change point alternatives.   

\section{Empirical Type One Error and Power Simulations}

Prior to analyzing the outcome of interest, we conduct simulations to (1.) examine the type one error rate, and (2.) determine the power and accuracy of our proposed methodology to detect a global change point in the mean functions of the response. 
These simulations examine the operating characteristics of our proposed supremum Wald test under various conditions. We continue to test 
\begin{align*}
H_0 &: \,  \delta_j=\Delta_j=0 ~~ \forall ~j, ~j=1,\dots,J\text{ (no change point)} \\
\text{vs.} ~~H_a &:  \,  \,  \delta_j \ne 0 \text{ and/or } \Delta_j \ne 0, ~ \text{for some} ~j,  ~j=1,\dots,J,  ~  \text{ (a change point at} ~q  \text{)}
\end{align*}
with $q \in Q$ (the set of possible change points specified by the researcher). The full and reduced models of these simulations are those of equations (\ref{FullModels}) and (\ref{NullModels}), respectively. We have additionally examined the scenario with standardized quadratic time (and standardized linear time) in the mean functions of the reduced and full models. We focused on standardized time, as opposed to untransformed time, to avoid collinearity between the two time terms. These simulations are omitted for brevity, though we note that we obtain similar results as those discussed in the following sections. In both sets of simulations we assume an autocorrelation structure that remains constant over the entire duration of the observational period, since the focus is on testing for the existence of a change point in the mean functions.

The outcome of interest is recorded for $60$ time periods, in five units, with adjacent correlation estimates smaller than $\phi = 0.1,$ and the set of possible change points equal to $\{25, \dots, 34\}.$ Thus, we chose parameters similar to these values for our simulations. We consider two values of the time series length, $T \in \{ 60, 120\},$ two values of the adjacent correlation, $\phi \in \{ 0.1, 0.6\},$ and three values for the total number of units, $J \in \{ 1, 3, 5\}.$ 
When the length of the time series is $T=60$, we allow the set of possible change points to be $Q_{60}=\{25, \dots, 34\},$ as with the patient satisfaction data. In this situation we conduct $10$ total tests, since there are 10 elements in $Q_{60}$.
When the length of the time series is $T=120$ we allow the set of possible change points to be $Q_{120}=\{50, \dots, 69\};$ a total of 20 tests are conducted for $Q_{120}$. 

We choose to compare two values of the time series length to illustrate the possible gain in efficiency longer time series provide with regards to power. We illustrate the gain that may come from doubling the length of the time series.  The length of the time series can be increased in two ways: (1) increase the observational period, say from 5 years to 10 years; and/or (2.)  increase the resolution of recordings, i.e., record patient satisfaction bi-monthly, as opposed to monthly. The two values of $\phi$ examined are larger adjacent correlation values than what we estimate for the patient satisfaction data. The largest unit-specific adjacent correlation estimates obtained for the patient satisfaction data (when information is not borrowed across units) is $0.09$, and so  $0.1$ is an upper bound for the adjacent correlation in our setting.
The value $\phi=0.6$ represents an upper bound for the correlation between repeated measurements in the literature. The estimated adjacent correlations for patient satisfaction are smaller than either $0.1$ and $0.6.$ Our simulation results are conservative because power decreases for ITS designs as the adjacent correlation increases. \cite{zhang2011} 

Importantly, we conduct type one error and power simulations for the supremum Wald test with one, three and five units. We examine the case with a single unit, $J=1,$ to illustrate the performance of our supremum Wald test in the traditional ITS analysis setting. We explore the value $J=3$ to depict the healthy gain in efficiency that borrowing information across a small number of units yields. Lastly, we consider $J=5$ because patient satisfaction is recorded at five units. 
Our aim is to highlight the improvement in power that borrowing information across units can provide.

\subsection{Empirical Type One Error for SWT}

We provide the empirical type one error rates when testing for the existence of a change point via the supremum Wald test. Four different scenarios are considered for one unit, three units, and five units.   
We generate $10,000$ time series for each scenario under the reduced model --- i.e.,  from one overall regime where there is no change point present in either the mean functions or stochastic processes.   
For the case when there is only one unit we set the mean function parameters to $\vec{\beta} = [65, 0.5]^{'}.$ When there are three units or five units, the mean function parameters vary slightly across individual units. 
The white noise standard deviation, $\sigma_w,$ is always set to 3.38, regardless of the number of units in the simulation. The value $\sigma_w = 3.38$ is approximately the average of the single unit estimates of the white noise standard deviation for patient satisfaction. 
The response standard deviation, $\sigma$, is 4.23 when $\phi=0.6$ and 3.40 when $\phi=0.1$ for all individual units. The mean function parameter values mimic results obtained from the patient satisfaction data.  

The empirical type one error rates for each scenario are provided in table \ref{MultiTOE}. As expected, the empirical type one error rate is smaller for the longer time series, and for smaller values of the adjacent correlation. Larger values of adjacent correlation imply a smaller number of effective independent statistical information.
The larger adjacent correlation quantity corresponds to higher type one error rates exclusively. In the scenario with the shorter time series and high adjacent correlation value, it is more difficult to control the family-wise type one error rate, even as we increase the number of units. The lack of type one error rate control in short time series with high correlation values is exacerbated in the simulations with quadratic time in the mean functions. In fact, for that particular setting the type one error rate becomes worse as the number of units increases. 
This is primarily attributable to the increased dependency in an already short time series that reduces the information in the time series. Because of this we recommend our proposed procedure when the length of the time series is at least 120 time points in cases with complex mean functions and/or hight correlation values. For all other scenarios considered, the type one error rate is well controlled, though slightly conservative because of the Benjamini-Hochberg multiplicity correction.

\begin{table}[!htbp]
	\begin{center}
		\begin{tabular}{||c|c|c|c|c|c|c|} 
			\hline
			Adjacent & \multicolumn{3}{c}{Time Series of Length 60} \vrule & \multicolumn{3}{c}{Time Series of Length 120} \vrule \vrule \\ \hline
			Correlation & One Unit & Three Units & Five Units & One Unit & Three Units & Five Units \\
			\hline \hline
			$\phi = 0.1$ & 0.0295 & 0.0291 & 0.0342 & 0.0274 & 0.0265 & 0.0263  \\  [0.5ex] 
			\hline
			$\phi = 0.6$ & 0.0460 & 0.0704 & 0.1003 & 0.0299 & 0.0318 & 0.0436 \\ \hline \hline
		\end{tabular}
		\caption{The empirical type one error rate: the proportion of iterations for which we rejected the null hypothesis of no change point. The larger adjacent correlation quantity corresponds to higher type one error rates exclusively. The type one error rates are reasonable for almost all of the scenarios, and stay reasonable as the number of units increases. However, it is slightly difficult to control the type one error rate at the desired $\alpha = 0.05$ level with the smaller time series and high adjacent correlation value. }
		\label{MultiTOE}
	\end{center}
\end{table}

\subsection{Empirical Power for SWT}

We conduct simulation-based power calculations when testing for the existence of a change point via the supremum Wald test. Time series are generated under the alternative model appropriate in our setting, i.e., generated with a global change point in the mean functions.
The change point is set at the middle of the time series; cases with the change point at the boundary or close to the boundary have been considered and yield similar, yet slightly less powerful, results. We focus on providing power as a function of the slope change.  Simulation-based power calculations with power as a function of the auto-regressive coefficient for ITS designs are provided in Zhang, Wagner and Ross-Degnan (2011). \cite{zhang2011}

Power is examined as a function of the slope change, with a change in baseline intercept ($\delta_j$) set to zero. Note, estimates of $\delta_j$ obtained from the patient satisfaction data are not statistically different from zero. The range of values for the change in slope, $\{0, 0.01, \dots, 0.24, 0.25, 0.30, \dots, 0.40, 0.45\}$, encompass the estimated quantities of the change in slope for the patient satisfaction data. Similarly to the type one error simulations, the white noise standard deviation is set to $3.38$ for all units, yielding a response standard deviation of $4.23$ when $\phi=0.6$ and $3.40$ when $\phi=0.1.$

\begin{figure}[!htbp]
	\centering 
		\includegraphics[scale=.16]{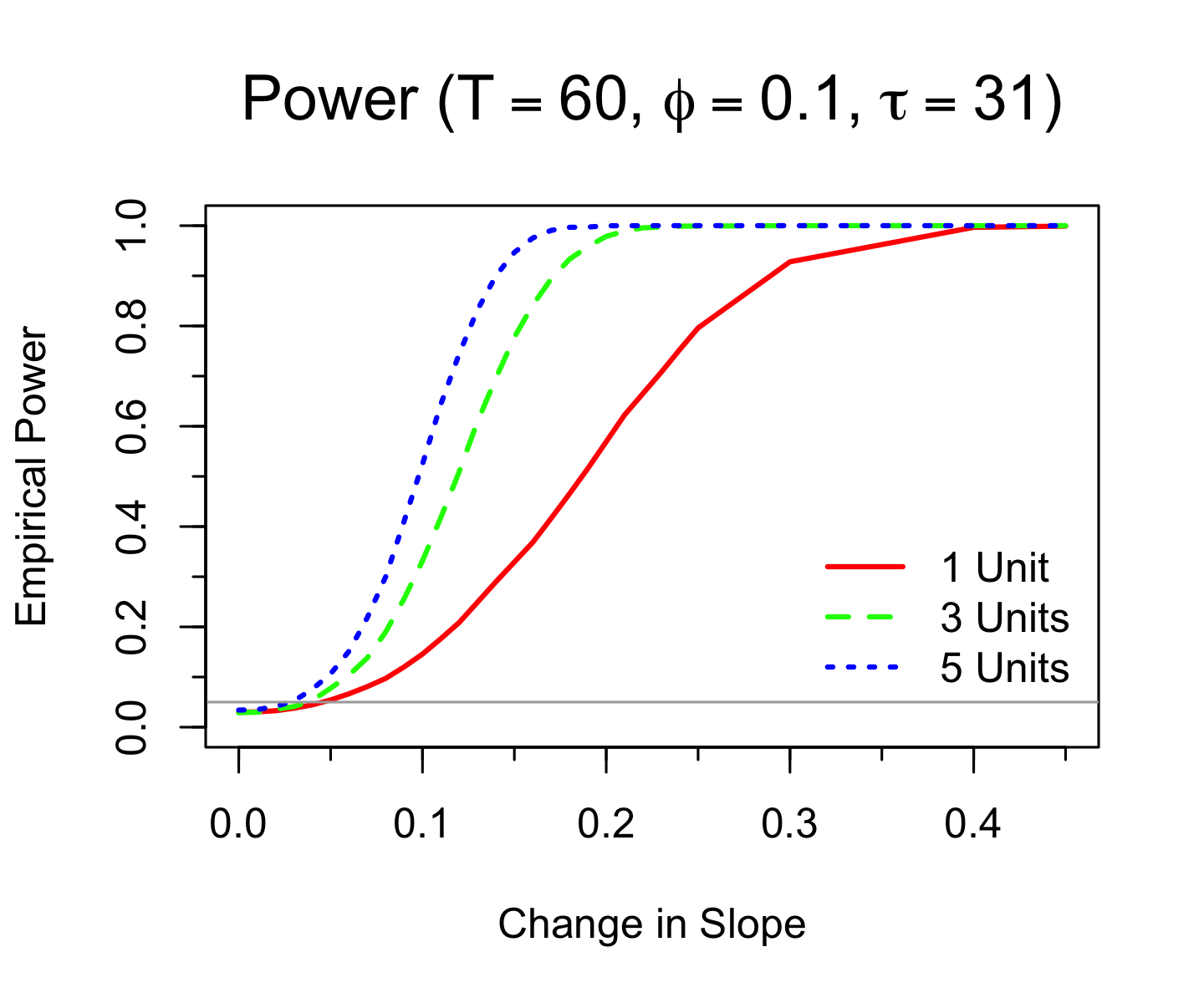}
		\includegraphics[scale=.16]{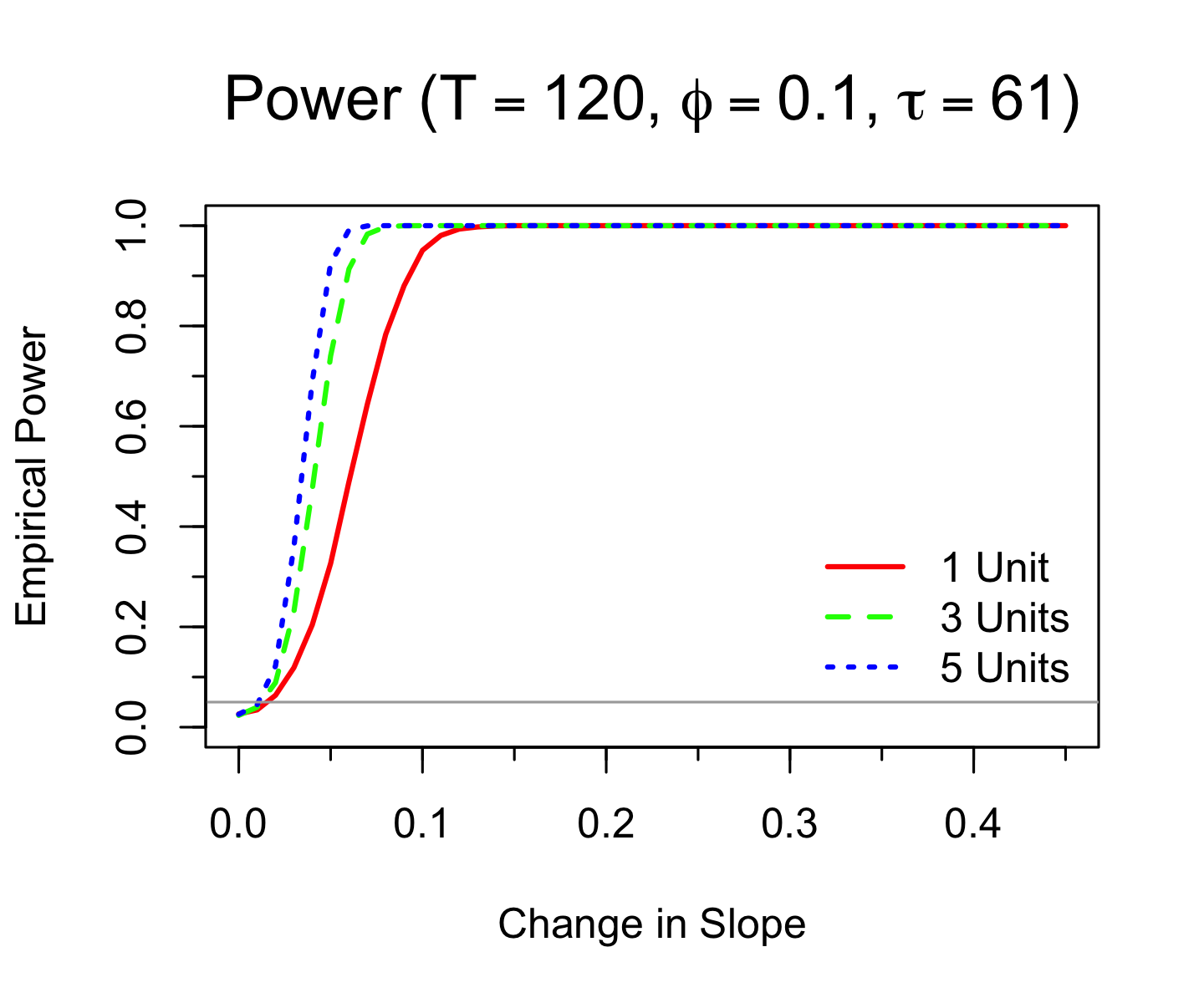}
		\includegraphics[scale=.16]{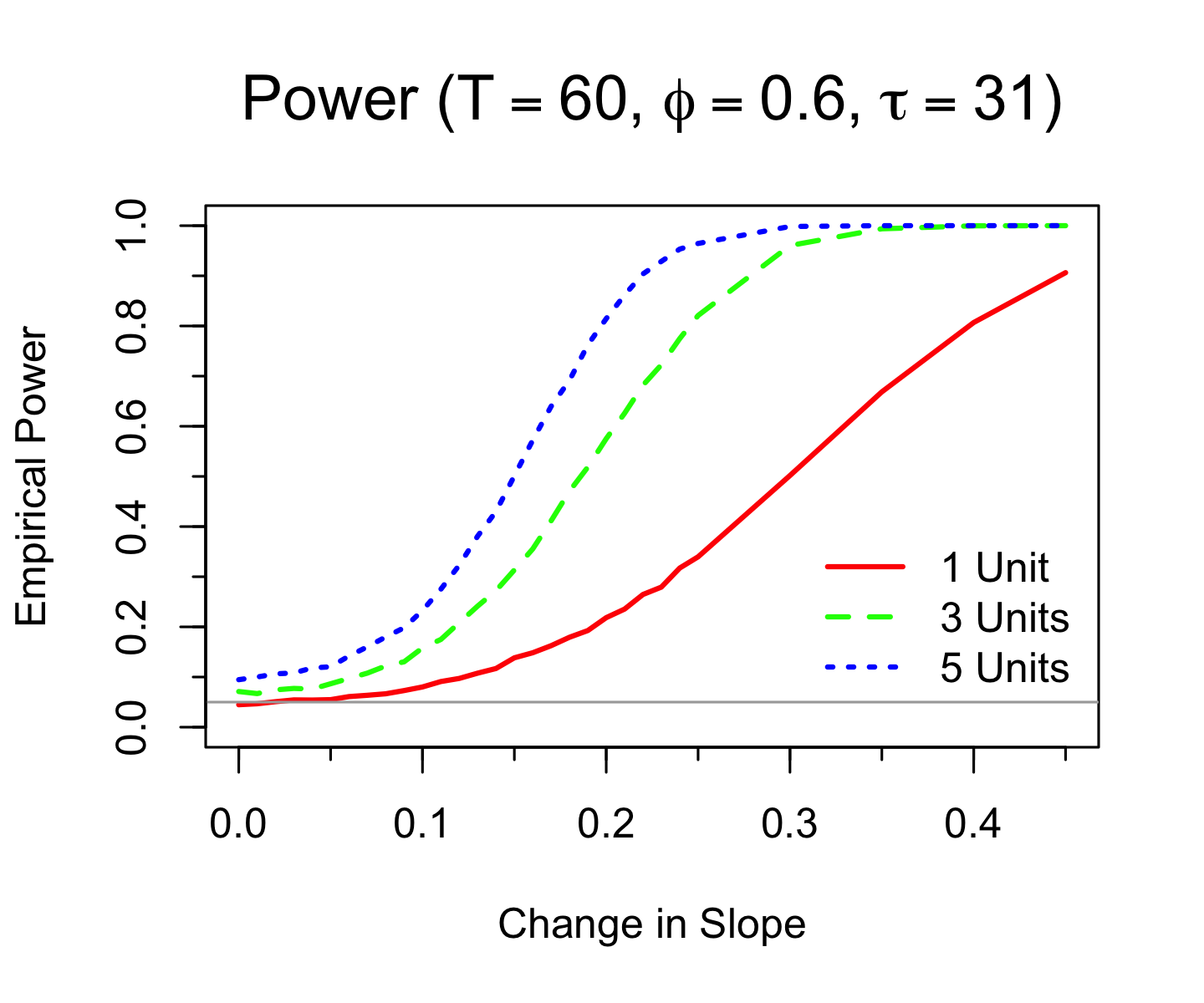}
		\includegraphics[scale=.16]{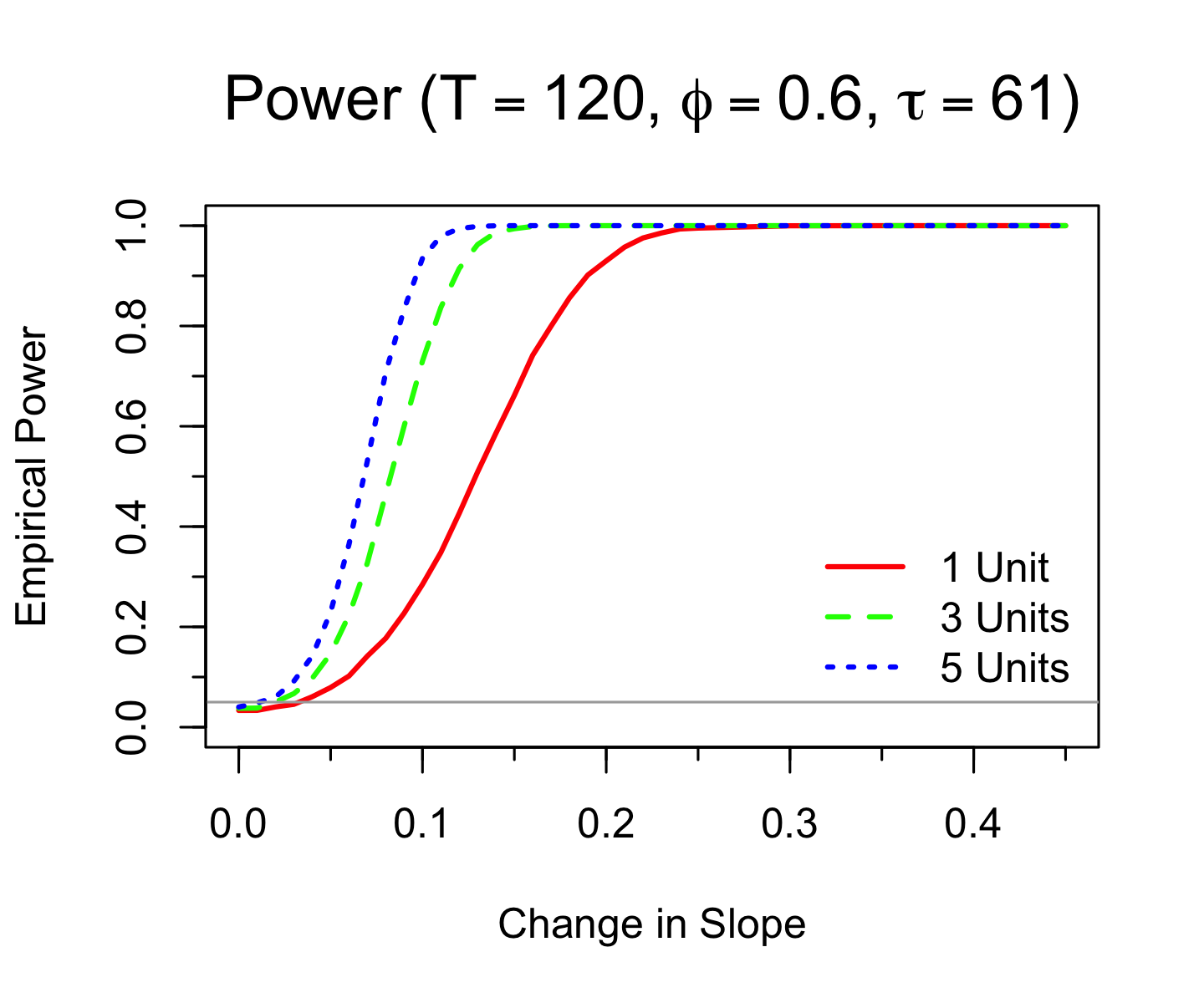}
	\caption{Empirical power, over $10,000$ iterations, for various number of units and for 4 regimes. The empirical power increases as the number of units and the length of time series increases, and the power increases as the adjacent correlation decreases.}
	\label{Power}
\end{figure}

Simulated power curves are provided in Figure \ref{Power}, with each subfigure corresponding to a separate data generation regime. As expected, power increases as the slope change and length of the time series increases, and power decreases for the larger adjacent correlation value.
Power is consistently higher for the larger number of units across the four scenarios, thereby illustrating that the supremum Wald test gains power as the number of units increases by borrowing information across units. Analyzing multiple time series data (or data from multiple hospital units) jointly, results in higher power. 

\subsubsection{Accurate Estimation of the Change Point}

\begin{figure}[!htbp]
	\centering 
		\includegraphics[scale=.16]{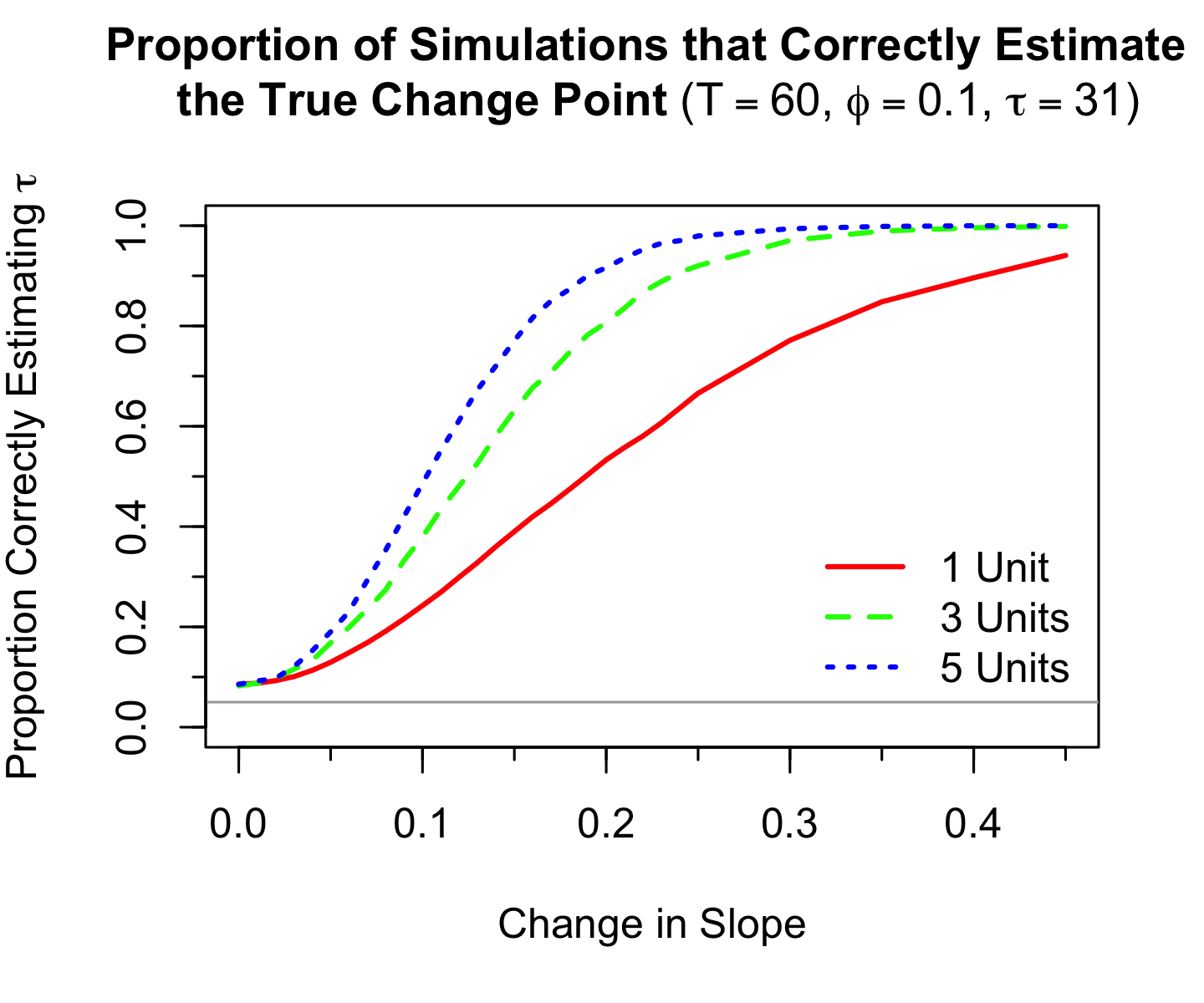}
		\includegraphics[scale=.16]{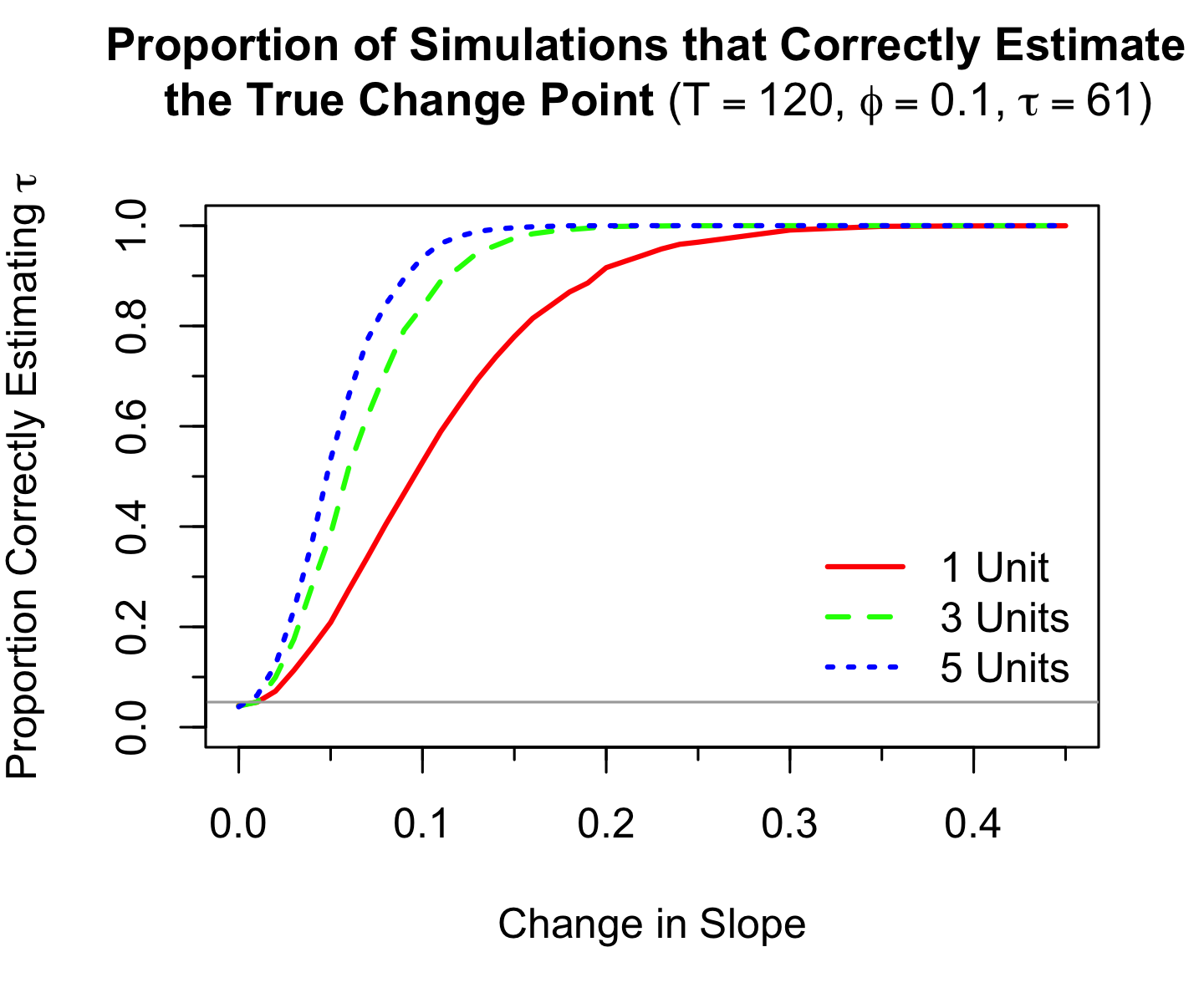}
		\includegraphics[scale=.16]{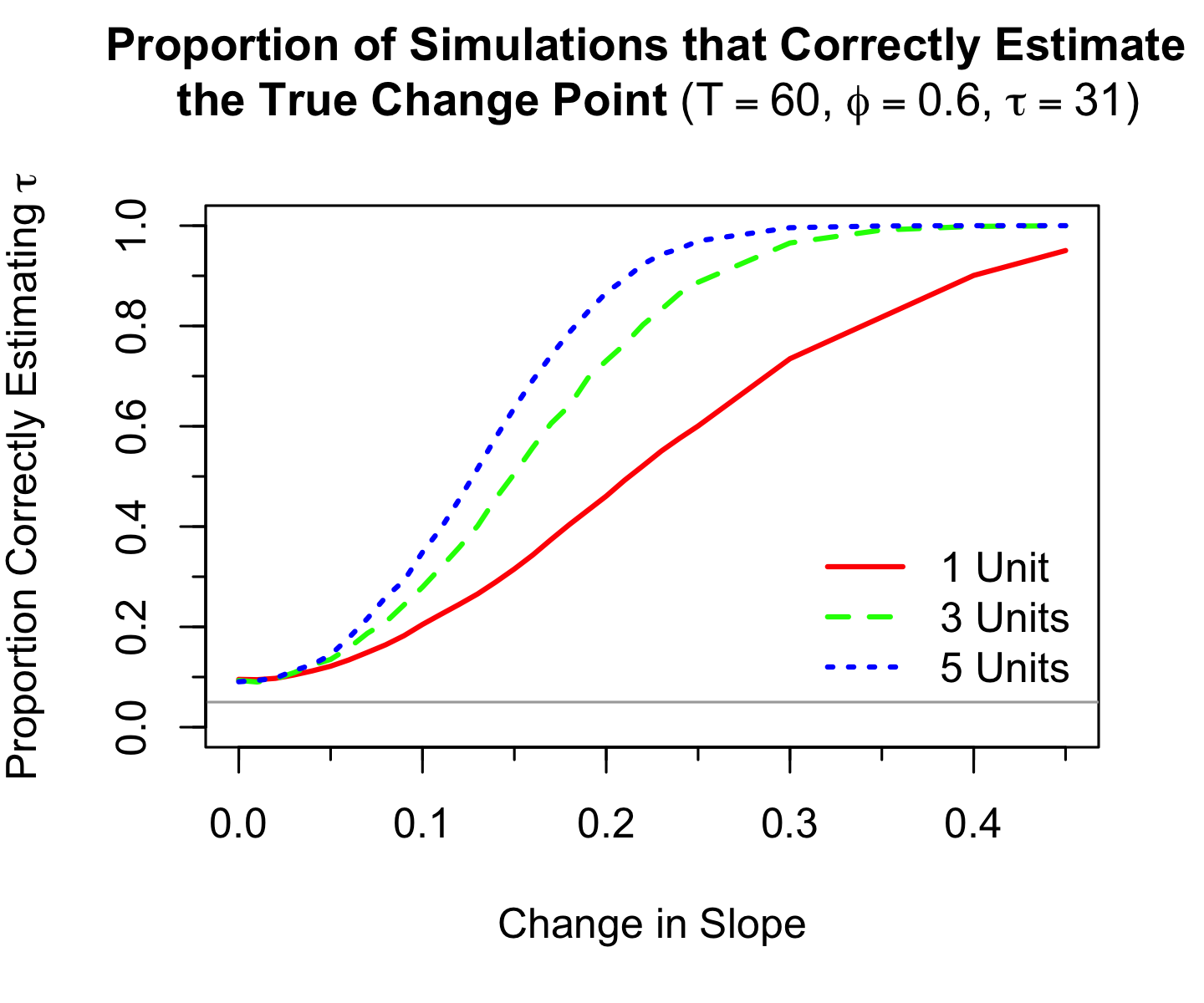}
		\includegraphics[scale=.16]{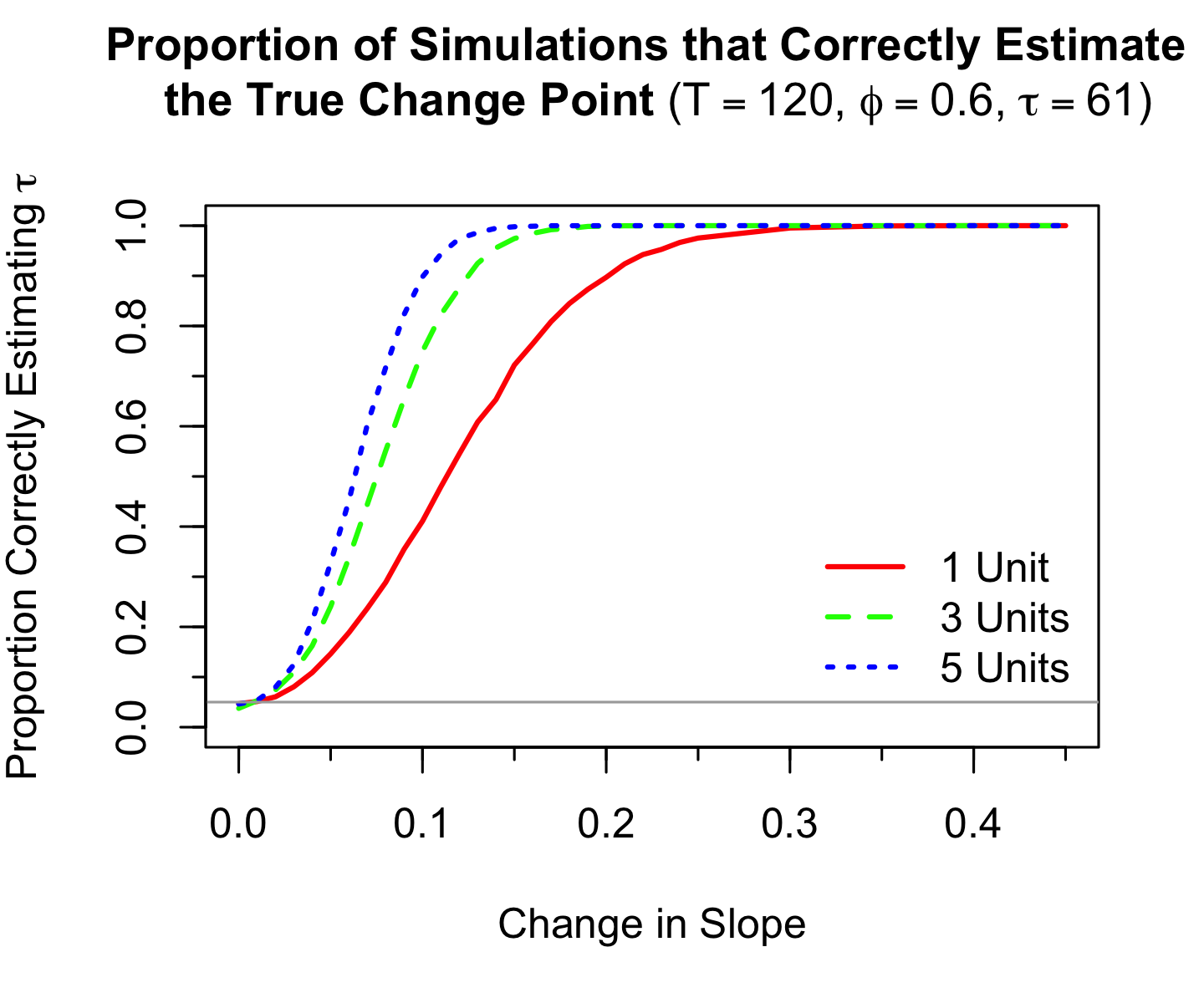}
\caption{The proportion of estimated change points exactly equal to the true change point, over $10,000$ iterations, for various number of units and for 4 regimes. Similar to the empirical power, the proportion of correctly estimated change points increases as the number of units and the length of time series increases.}
\label{coverage}
\end{figure}

The power simulation results, provided in Figure \ref{Power}, suggest that the supremum Wald test has reasonable power to detect an existing global change point, and that power increases as the number of units increase.  We are not simply interested in power by itself. We are also interested in whether R-MITS will provide the correct global change point estimate when our supremum Wald test concludes that a change point exists. 
Figure \ref{coverage} illustrates the proportion of simulations that correctly estimate the true change point as a function of the slope change for one, three and five units. Similar to the empirical power, the proportion of correctly estimated change points increases as the number of units and the length of the time series increases. We also calculated the proportion of simulations that exactly estimate the true change point for change points not in the middle of the time series --- i.e., with a change point on the boundary or near the boundary --- and obtained comparable results. 

\section{Multi-Unit Analysis of the CNL Intervention}

We assess the impact of the Clinical Nurse Leader (CNL) integrated care delivery intervention on average patient satisfaction at five hospital units. 
Average patient satisfaction is the mean of patient satisfaction survey scores for seven indicators, shown for the Stroke and Surgical units in Figure \ref{UnitTS}. The seven patient satisfaction indicators are: effective nurse communication, nurses treated me with courtesy/ respect, responsiveness of hospital staff, effective physician communication, staff did everything to help control your pain, effective communication about medicines, and  discharge information provided.  
We refer to the average patient satisfaction scores simply as patient satisfaction. 

We are interested in estimating the time lag (or delay) between the onset of the intervention and the effect on patient satisfaction. 
In practice, the change point may occur either \textit{before} or \textit{after}
the official intervention time. An intervention intended to improve care quality requiring a training over several months or weeks (such as the CNL intervention)
may already produce a change in the outcome, even before the official start of intervention, if the trainees execute their training as they learn. 

Inference on the global change point and time lag between the onset of the intervention and the intervention's effect is included in Table \ref{CP}. Table \ref{CP} provides the (a.) global change point estimate (b.) supremum Wald test p-value, (c.) time point of Clinical Nurse Leader integration into their respective hospital units, (d.) formal intervention implementation time, and (e.) lag between formal intervention implementation and estimated change point. 
The supremum Wald test concludes that a change point exists over the set of possible change points for patient satisfaction in at least one of the units at the $\alpha=0.05$ level.
The p-value associated with the test for the existence of a change point is $0.003,$ which is less than the respective Benjamini-Hochberg corrected critical value of $0.005$. 
R-MITS estimates a preemptive effect of the CNL integrated care delivery intervention on patient satisfaction. The global change point is estimated to occur on May 2010, while the formal intervention implementation occurs on July 2010. Estimating an anticipatory effect (from the expected and a priori specified change point) is not feasible with standard segmented time series regression. Segmented Regression methodology requires clearly separated pre- and post-intervention phases, often with an assumed change point greater than or equal to the formal intervention implementation time point. 

\begin{table}[!htbp] 
	\begin{center}
		\begin{tabular}{||c| c c| c |c|} 
			\hline
			CNL & Estimated &  P-value$^{+}$ & Formal Intervention  &  Lag** \\
			Introduction* & Change Point $\widehat{\tau}$ &   & Implementation &  \\ [0.5ex] 
			\hline\hline
			January 2010 &   May 2010 & 0.0003 &  July 2010  &  -2   \\
			\hline
		\end{tabular}
	\end{center}
	\caption{Provides the estimated global change point, its p-value, the month Clinical Nurse Leaders were integrated into their respective units, the formal intervention time point, and the intervention lag. The Benjamini-Hochberg corrected p-value cut-off is 0.005. We conclude that there is a change point in patient satisfaction at the $\alpha= 0.05$ level, because $0.0003 \, < \,0.005.$ \,
    $^{+}$ The p-value for the supremum Wald test, i.e., the p-value for the existence of a change point. \,
	* All clinical nurse leaders were integrated into their respective hospital units on January 2010. \,
	** The intervention lag is the difference between the estimated change point and the formal intervention introduction time point.}
 
	\label{CP}
\end{table}

Although the CNL integrated care delivery is officially implemented on July 2010, it was unofficially being practiced prior to July 2010. Nurses put into practice the new concepts they learned from their `training'. 
It is completely realistic that many of the CNLs implemented their training prior to July 2010, particularly, if they believed it would be beneficial. 
Thus, the anticipatory effect of the CNL integrated care delivery intervention (of two 2 months, provided in the `Lag' column of Table \ref{CP}) is consistent with the integration of the CNLs on January 2010. 
In fact, the estimated global change point for patient satisfaction occurs four months after the Clinical Nurse Leaders introduction into their respective units. The CNL care delivery intervention requires a restructuring of patient care and care delivery, likely to manifest itself to patients after a time lag from the CNLs introduction. This time lag and the behavioral component of the intervention may explain why the global change point occurs four months after the CNLs integration into the hospital units and two months prior to the formal intervention time point.

Estimates of the R-MITS mean function parameters are provided in Tables \ref{PS} and \ref{PS2}, and estimates of the stochastic process parameters are included in Table \ref{AR}. 
Estimates and 95\% confidence intervals of the two standardized effect sizes used in the health care literature, change in level and change in trend/slope, \cite{EPOC:2015ww} are provided in table \ref{PS2}. The level and trend change are not statistically significant for any unit.
The estimated level change tends to be positive for the majority of hospital units, indicating an initial drop of the outcome level, as in Figure \ref{CIL}. This may be due to the adjustment period associated with the intervention. Moreover, it may occur as an artifact of the regression itself, particularly for a bounded outcome such as patient satisfaction. 

\begin{table}[!htbp] 
	\begin{center}
		\begin{tabular}{|c||c|c|c||c|c|c|} 
			\hline
			Hospital Unit & \multicolumn{3}{c}{Intercept Pre-Change Point \, $\widehat{\beta}_{j0}$} \vrule & \multicolumn{3}{c}{Slope Pre-Change Point \, $\widehat{\beta}_{j1}$ } \vrule \\ \hline
			& Estimate & 95\% CI & p-val & Estimate & 95\% CI & p-val \\
		\hline\hline
		Stroke & 64.32 & ( 46.34, 82.31 ) & 0 & 0.56 & ( -0.52, 1.64 ) & 0.3  \\ \hline 
		Surgical & 72.8 & ( 47.72, 97.88 ) & 0 & 0.36 & ( -1.05, 1.77 ) & 0.61  \\ \hline 
		Cardiac & 64.17 & ( 37.08, 91.27 ) & 0 & 0.31 & ( -1.3, 1.92 ) & 0.7  \\ \hline 
		Medical Surgical & 70.19 & ( 41.77, 98.61 ) & 0 & 0.19 & ( -1.53, 1.91 ) & 0.83  \\ \hline 
		Mother/baby & 77.1 & ( 63.1, 91.09 ) & 0 & 0.28 & ( -0.58, 1.15 ) & 0.52  \\ \hline
			\hline
		\end{tabular}
	\end{center}
	\caption{The unit specific pre-change point intercepts and slopes. }
	\label{PS}
\end{table}	

\begin{table}[!htbp] 
	\begin{center}
		\begin{tabular}{|c||c|c|c||c|c|c|} 
			\hline
			Hospital Unit & \multicolumn{3}{c}{Change in Level \, $ - \widehat{\delta}_j - \widehat{\Delta}_j \widehat{\tau}$ } \vrule & \multicolumn{3}{c}{Change in Slope \, $\widehat{\Delta}_j$} \vrule \\
			\hline
			& Estimate & 95\% CI & p-val & Estimate & 95\% CI & p-val \\
			\hline\hline
		Stroke & 6.91 & ( -14.65, 28.46 ) & 0.52 & -0.35 & ( -1.59, 0.89 ) & 0.58 \\ \hline 
		Surgical & 6.17 & ( -20.22, 32.56 ) & 0.64 & -0.21 & ( -1.87, 1.45 ) & 0.8 \\ \hline 
		Cardiac & -0.15 & ( -34.36, 34.06 ) & 0.99 & -0.22 & ( -2.25, 1.82 ) & 0.83 \\ \hline 
		Medical Surgical & 0.3 & ( -40.57, 41.18 ) & 0.99 & -0.14 & ( -2.53, 2.24 ) & 0.9 \\ \hline 
		Mother/baby & 3.73 & ( -22.1, 29.56 ) & 0.77 & -0.25 & ( -1.72, 1.23 ) & 0.74 \\ \hline 
			\hline 
		\end{tabular}
	\end{center}
	\caption{The unit specific change in levels and change in slopes.   }
	\label{PS2}
\end{table}	

\begin{figure}[!htbp] 
	\centering
		\includegraphics[scale=.125]{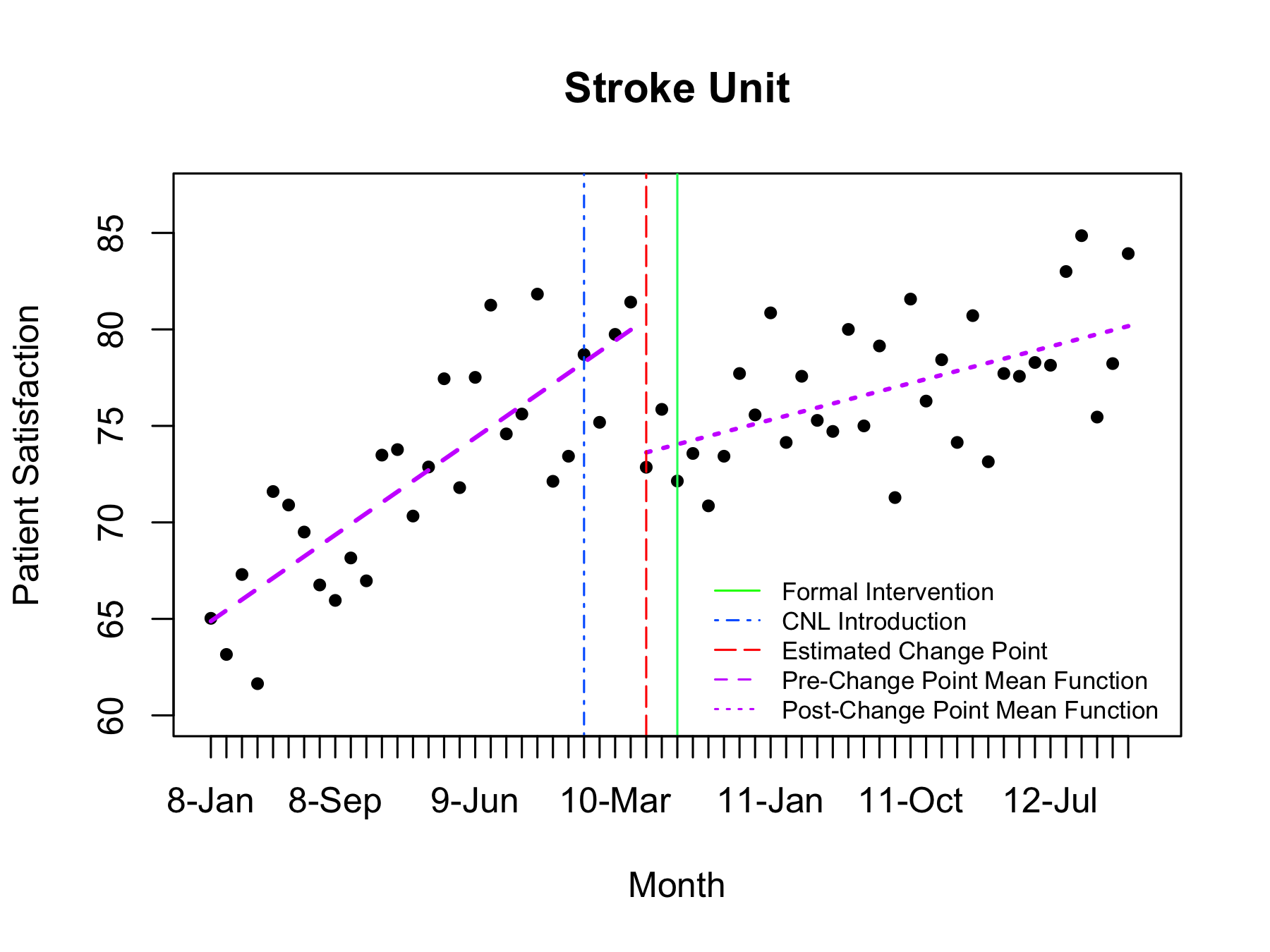}
		\includegraphics[scale=.125]{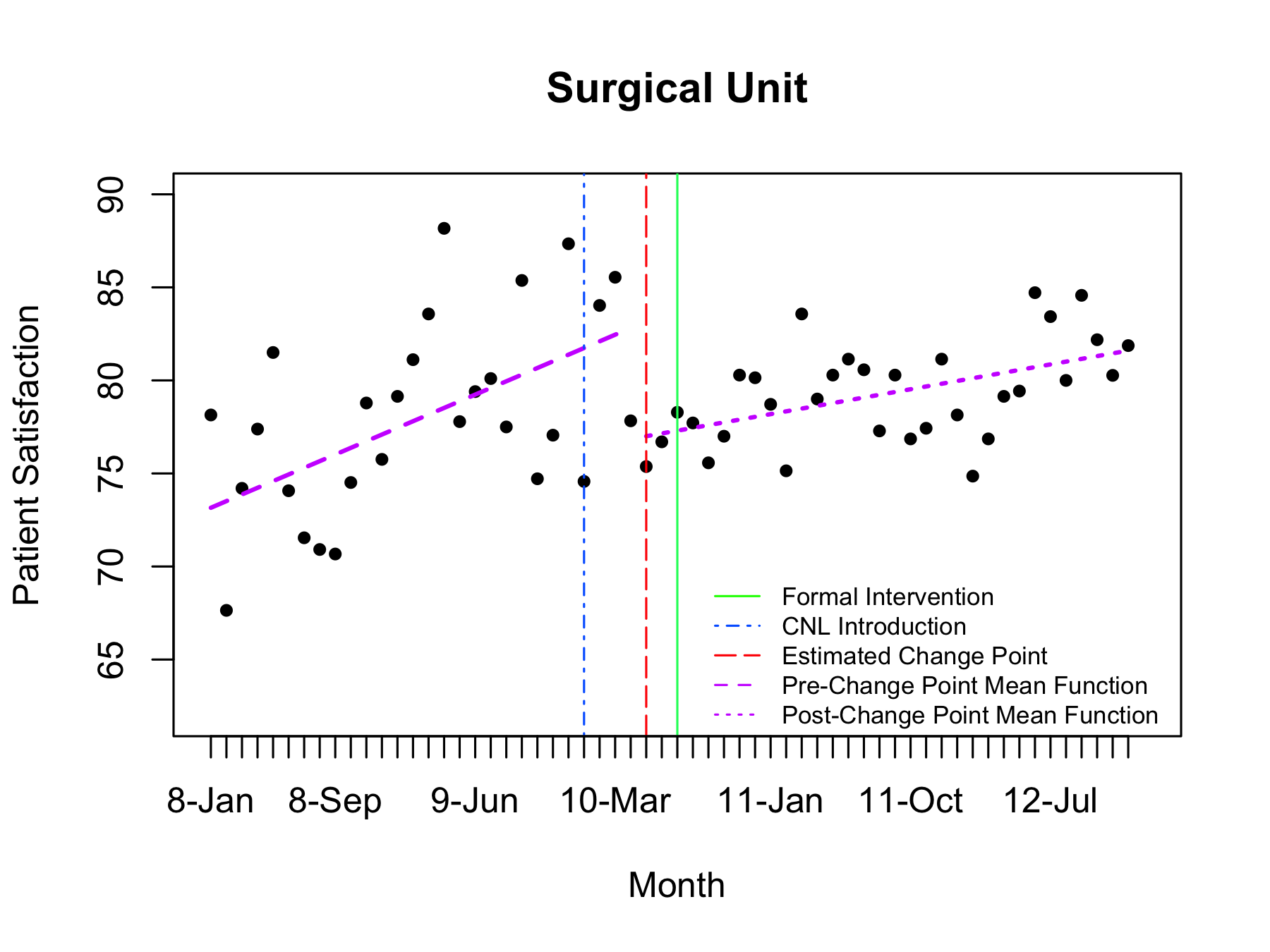}
	\includegraphics[scale=.125]{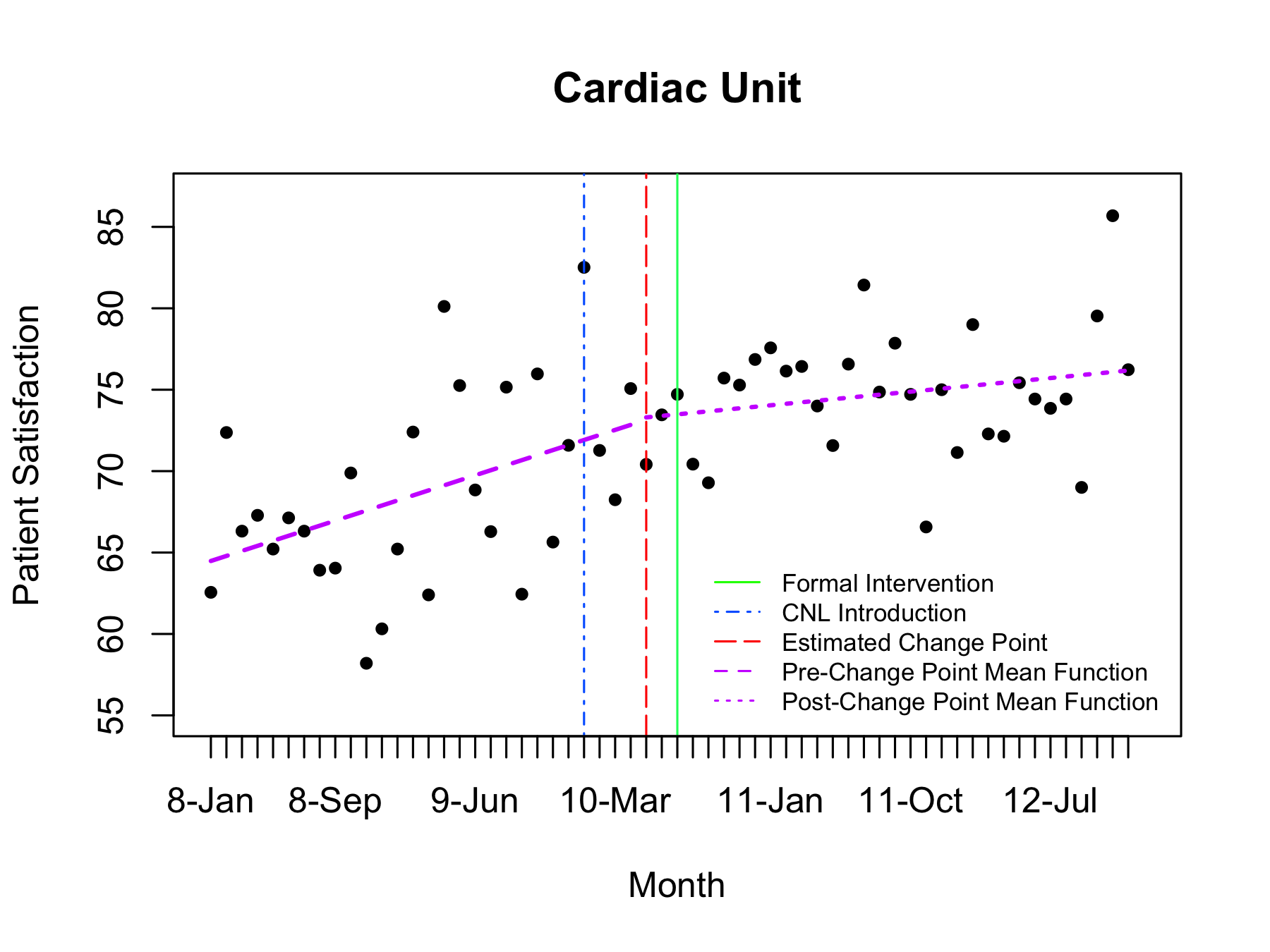}
\includegraphics[scale=.125]{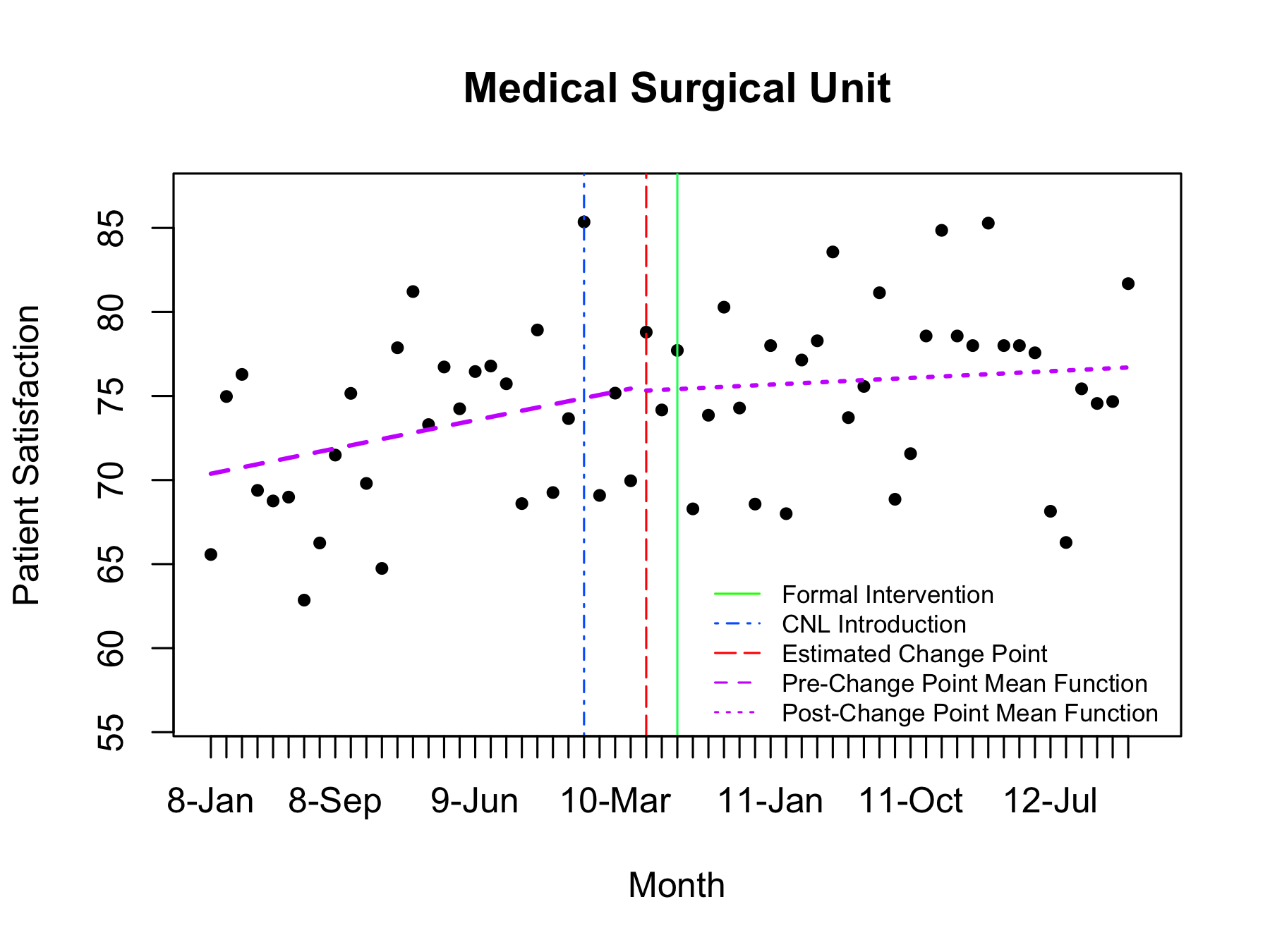}
		\includegraphics[scale=.125]{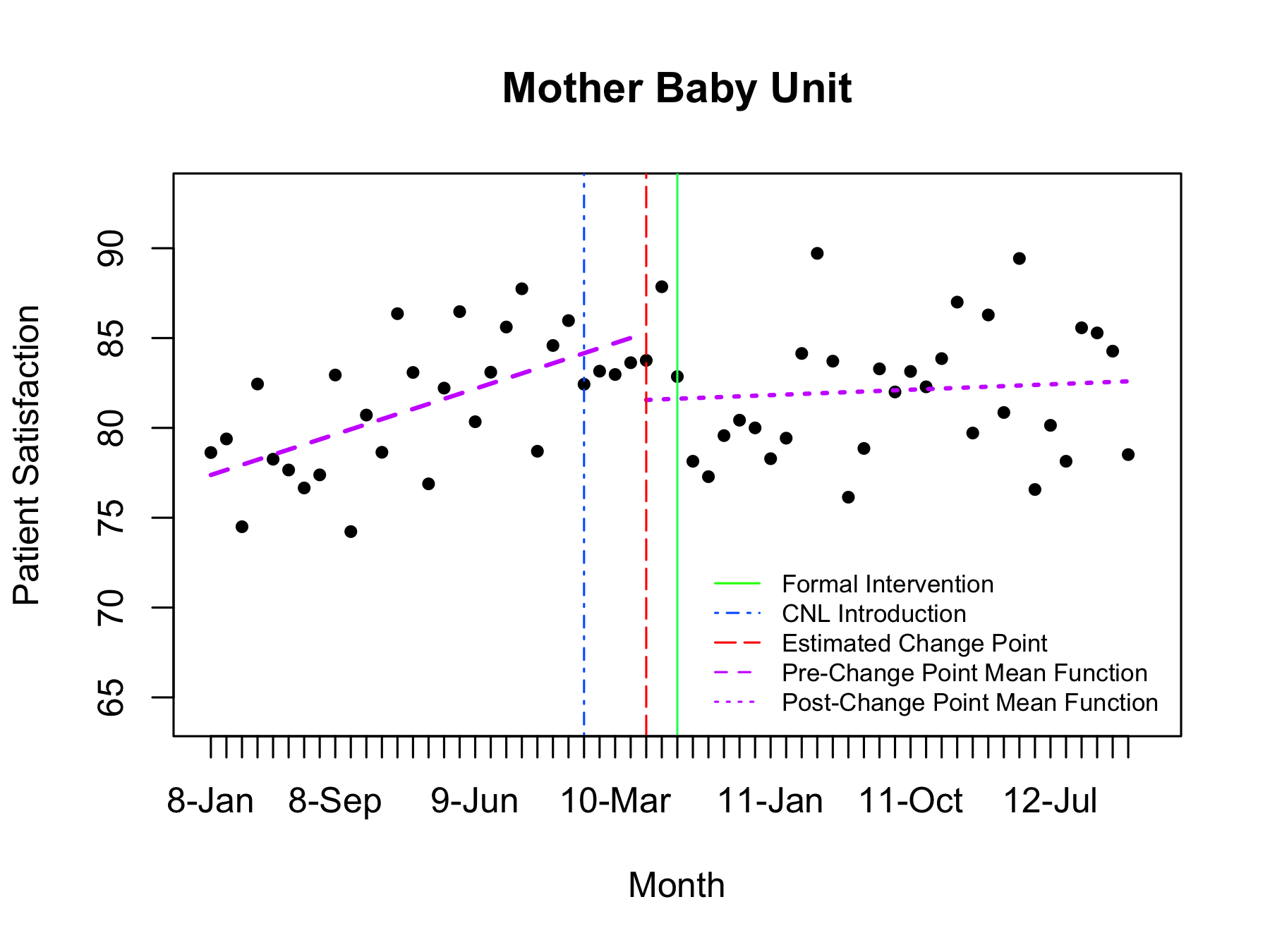}
	\caption{Plots the time series of observed average patient satisfaction for all hospital units, along with the estimated change point, estimated mean functions, and formal intervention time.}
	\label{Fitted1}
\end{figure}

Trend (slope) change is negative for patient satisfaction, suggesting a decrease in the slope of patient satisfaction post-change point. Due to the nature of patient satisfaction as a percentage --- and thus as a bounded outcome --- the change in slope must be interpreted with caution.
Patient satisfaction cannot continue to grow at a rapid rate because the mean patient satisfaction function at the estimated change point is already relatively close to 100, the maximum patient satisfaction value. This is evident in Figure \ref{Fitted1}, in which the estimated mean functions for all hospital units are plotted, particularly for the Stroke, Surgical, and Mother/Baby units. 

The estimated volatility of patient satisfaction, given by the `Standard Deviation' column of Table \ref{AR}, is smaller post-change point for 3 out of the 5 units, and the adjacent correlation is larger post-intervention in 4 out of the 5 units. The Medical Surgical and Mother/baby units estimated standard deviations increase post-estimated change point, increasing from 4.84 and 2.97 to 5.06 and 3.72, respectively; while in the Stroke, Surgical, and Cardiac units the estimated standard deviation decreases from 3.15, 4.37, and 5.28 to 3.01, 2.35, and 3.76, respectively. After the estimated change point, the patient satisfaction scores are observed to be less volatile for the Stroke, Surgical, and Cardiac units, and hence may be more predictable.  The adjacent correlation estimates mainly move from negative to positive post-estimated change point, indicating a more stationary patient satisfaction score post-intervention. These are positive results of the CNL intervention. It is important for hospitals to have patients that are generally satisfied over patients who range from extremely satisfied to extremely dissatisfied. Patient satisfaction scores that are more dependent, closely related, and less volatile result in a more predictable outcome. 

\begin{table}[!htbp]
	\begin{center}
	\begin{tabular}{||c|c|c|c|c||}
		\hline
		Hospital Unit & \multicolumn{2}{c}{Pre-Change Point} \vrule & \multicolumn{2}{c}{Post-Change Point} \vrule \\ \hline
		 & Adjacent  & Standard & Adjacent & Standard \\
		 & Correlation  & Deviation  & Correlation  & Deviation  \\
		 & $\widehat{\phi}_{j1}$ & $\widehat{\sigma}_{j1}$ &  $\widehat{\phi}_{j2}$ & $\widehat{\sigma}_{j2}$ \\
		\hline \hline
		Stroke & -0.06 &  3.15 & -0.35  & 3.01 \\
		\hline 
		Surgical & -0.02 &  4.37 &0.19  & 2.35 \\
		\hline
		Cardiac & -0.16 & 5.28  & 0.10 & 3.76 \\ 
		\hline
		Medical Surgical & -0.03 & 4.84 &  0.09   & 5.06 \\
		\hline
		Mother/baby & -0.27 &  2.97 & 0.08 & 3.72 \\
		\hline
	\end{tabular}
	\end{center}
	\caption{Estimates of the stochastic component parameters: the adjacent correlations and response standard deviations pre- and post-change points. All the adjacent correlations are relatively small, and tend to switch from negative to positive post-intervention. The response standard deviations tend to decrease post-intervention.  }
	\label{AR}
\end{table}	

\subsection{Doubly Robust ITS}

\begin{figure}[H] 
		\includegraphics[scale=.126]{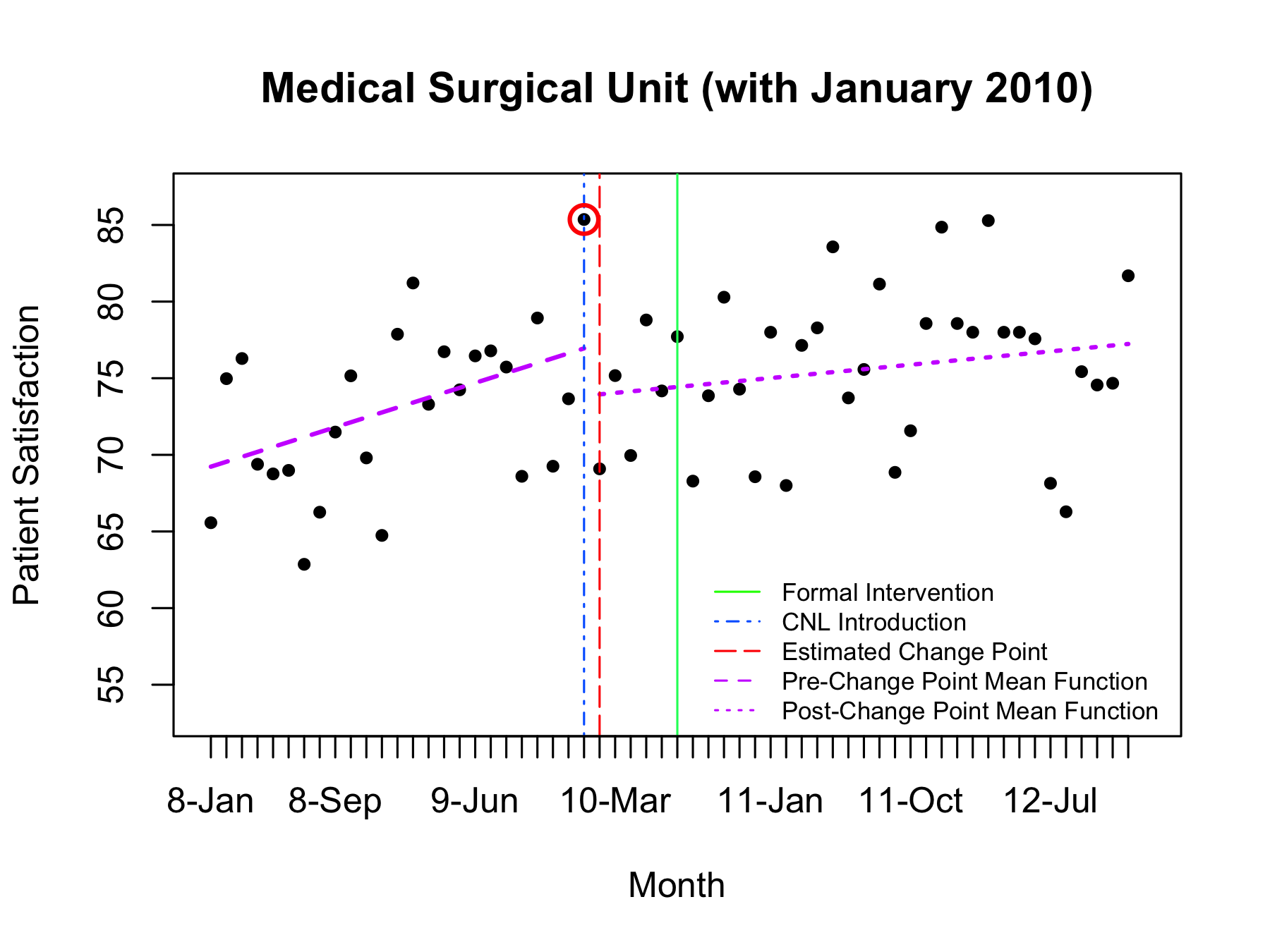}
	\includegraphics[scale=.126]{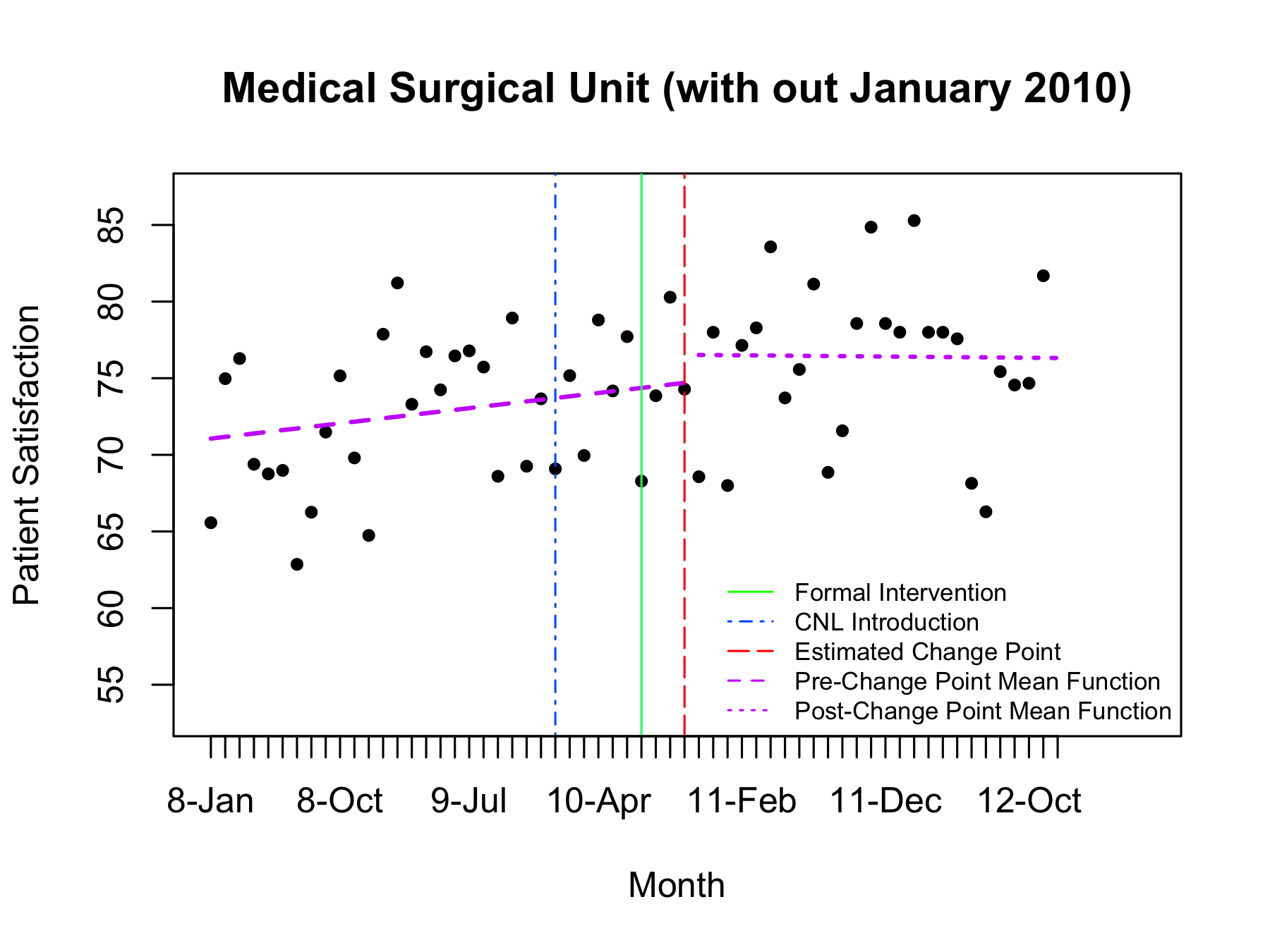}
		\includegraphics[scale=.126]{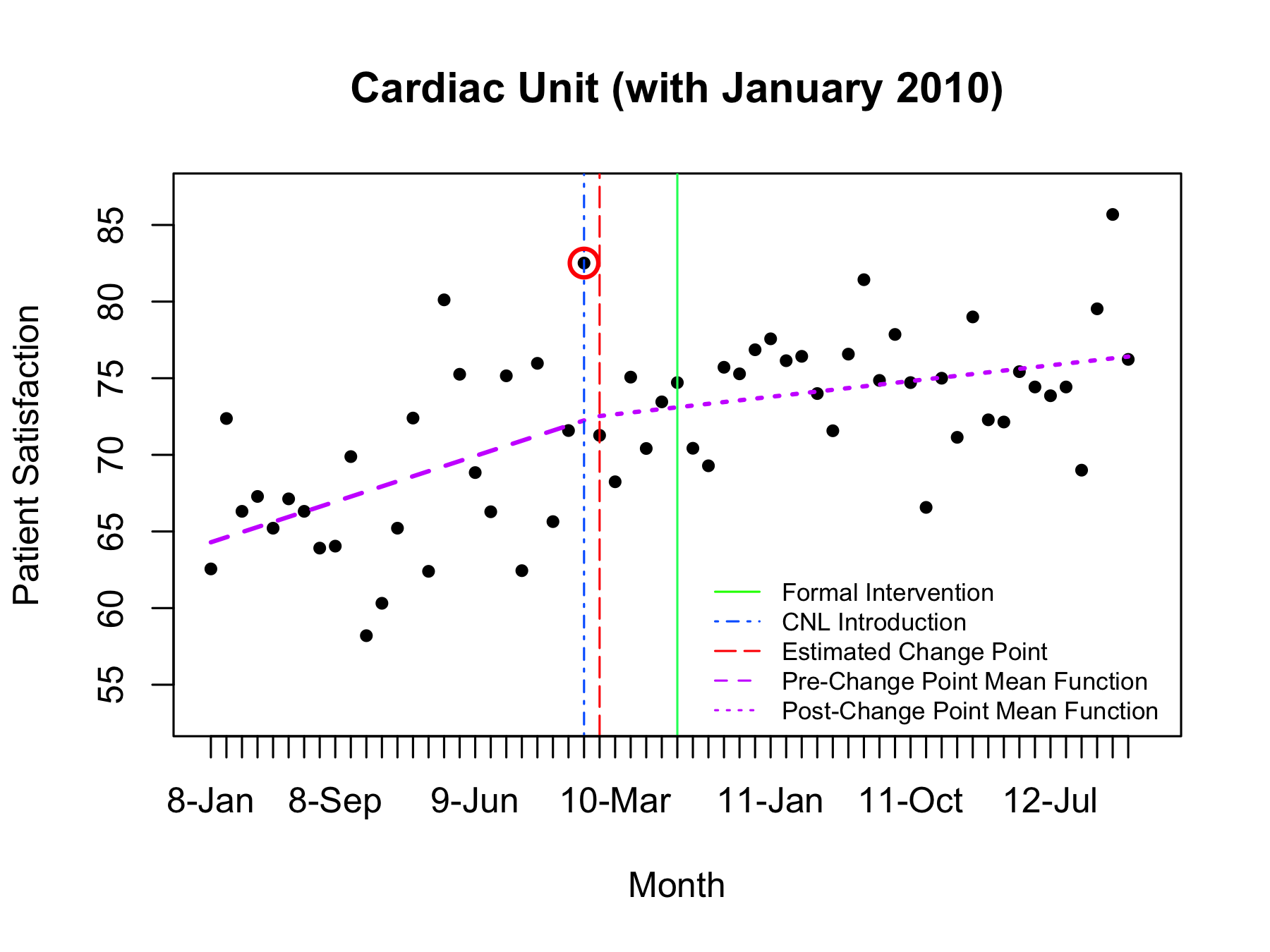}
\includegraphics[scale=.126]{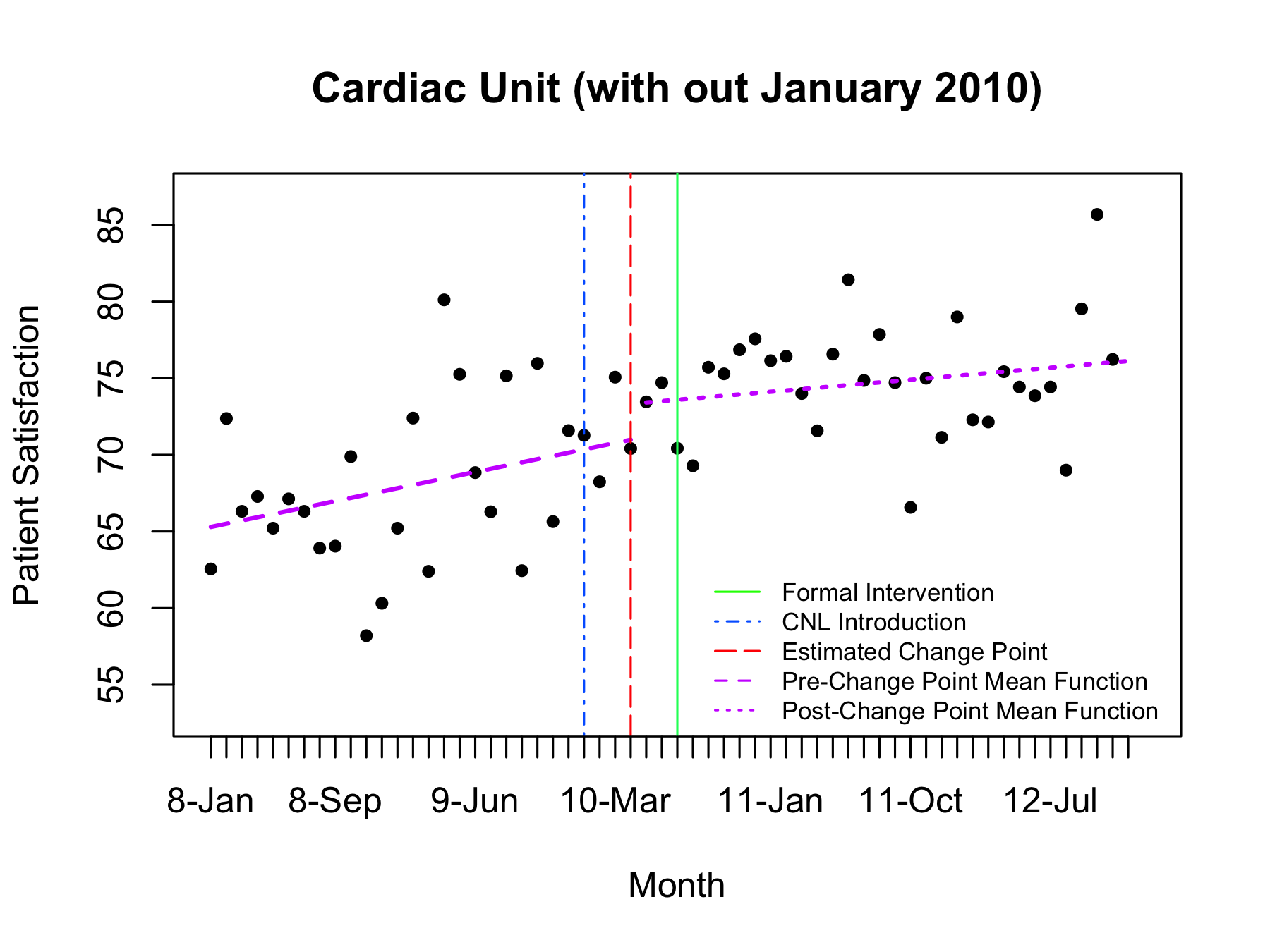}
	\caption{Plots the time series of observed average patient satisfaction, along with the estimated change point, estimated mean functions, and formal intervention time for the Medical Surgery and cardiac units with and without observation $t=25$ (January 2010), obtained by using Robust-ITS to conduct the unit specific analyses. \emph{ Note, the analysis with $t=25$ is on the left and the analysis without $t=25$ is on the right.}}
    \label{influential}
\end{figure}

R-MITS pools information across units to estimate a global change point, thereby increasing efficiency and reducing the impact of misleading influential points. 
Reducing the effect of influential points is desirable in our patient satisfaction data, for which the change point search space consists of only a few time points.
We illustrate the gravity of influential points on the estimated change point for the single unit analyses of patient satisfaction at the Medical Surgery and Cardiac units.
To model patient satisfaction for a single unit we implement the Robust-ITS model. 
The estimated mean functions and change point estimates are included in Figure \ref{influential} for two cases. The plots on the left of Figure \ref{influential} correspond to single unit analyses including all observations, while the plots on the right pertain to the single unit analyses without observation $t=25$ (January 2010).
When all the observations are included,
Robust-ITS estimates the change point to be February 2010 for both the Medical Surgical and Cardiac units. However, for the analyses without January 2010, the estimated change points are October 2010 and April 2010 for the Medical Surgical and Cardiac units, respectively.
One single time point has the ability to perturb the estimated change point by six months in the Medical Surgical unit and by two months in the Cardiac unit. 
Our proposed R-MITS model guards against these influential points by borrowing information across hospital units. Pooling data across hospital units in the estimation of a global change point automatically reduces the impact of spurious influential points, resulting in robust mean function estimates. 

\section{Conclusion and Future Work}

Our proposed R-MITS model is appropriate for multiple time series, able to estimate a global change point rather than assume it a priori, and can model differences in both the mean functions and stochastic components. 
R-MITS borrows information across units to estimate a global change point and to estimate the mean functions and stochastic processes separately for each unit. 
The proposed model does not assume that the impact of the global change point on the outcome is equivalent for all units. That is, although R-MITS borrows information across units to estimate an over-all-unit change point, the level change and trend change are allowed to vary for each unit. R-MITS further allows the autocorrelation and variability during pre- and during post-intervention to differ across units.

Importantly, our proposed supremum Wald test formally tests for the \emph{existence} of a change point in at least one of the mean functions, rather than merely assuming and requiring a change. Now researchers will be able to formally test whether an intervention is associated with a change in the mean functions of a health outcome. Erroneous inference regarding the response's mean functions may result from incorrectly assuming --- both the existence and placement --- of the change point. Assuming a change point when no change point truly exists forces the estimation of an artificial change. 
Our supremum Wald test will test for the existence of a change in the response over a pre-determined set of possible change points. As demonstrated by our simulation studies the operating characteristics of R-MITS and the supremum Wald test are well behaved with regards to power and type one error. Moreover, the empirical power of the supremum Wald test and accuracy of the change point estimates --- and so the accuracy of the estimated time delay between an intervention and the intervention's effect on an outcome --- increase as the number of units increases.

The R-MITS model and the supremum Wald test provide researchers with insight to re-address hypothesis generation for future study design. The methodology better informs researchers of the likely lag that may be realistic for a similar intervention. 
We note that in our application example nurses finishing their masters thesis project (in a program that trained them to implement the CNL intervention) were introduced into their respective hospital units six months prior to the formal intervention. The nurses integration potentially changed practice as soon as they were introduced. In fact, the estimated change point occurs between the introduction of the nurses to the hospital unit and the formal intervention.
A primary utility of R-MITS is that through exploration of the change point we are able to observe this and provide direction for future study planning.

R-MITS estimates the global change point via a grid search over a pre-determined set of possible change points. Researchers must specify the set of possible change points with care since, as with traditional ITS designs, we must be cautious of competing intervention effects. The set of possible change points must adequately capture the time points during which the intervention of interest plausibly impacted the outcome, yet simultaneously exclude time periods affected by another intervention. This is to avoid the risk of competing interventions. Parsing out the effect of competing interventions is a concern in general with ITS designs. Ideally, the entire observational period (both the pre- and post-intervention phases) of an ITS design should be solely affected by the intervention of interest. Although theoretically simple, in practice this requires careful consideration and expertise.

Identification of a change point via our proposed procedure relies upon detection of a difference in either the mean level of the response and/or the slope of the response, comparing the pre- and post-intervention effect periods across units. As such, if no change point in the time-series truly exists this would indicate that there is no difference in the mean function of the response over time. 
Most researchers would consider this absence of a difference in the mean function to be the absence of an intervention effect. One could argue that if the pre-intervention slope were positive (indicating improvement in outcomes) and if the slope remained constant during the post intervention, then this could have been solely attributable to the intervention. In this case the counterfactual may have revealed a decline (or an increase) in the slope if the intervention had not been instituted. Of course, such a counterfactual could never be observed in practice but certainly should be considered in theory.

Currently the supremum Wald test focuses on changes solely in the mean functions. 
We are currently working on extensions of the supremum Wald test to accurately detect changes is both the mean functions and stochastic components, to better handle the nuances of the autocorrelation structures across units. In addition, we are considering more efficient multiple testing corrections that utilize information obtained in the autocorrelation structure. We plan on developing the theory needed for our supremum Wald test (and its future derivatives) to guarantee consistency.

It is paramount to note that the current status of the R-MITS model is for continuous-valued outcomes only. We will soon expand this class of models to handle counts and rates data (e.g., infection rates, counts of accidental falls, etc). 
Lastly, R-MITS does not provide inference on the overall population of hospital units --- the population of hospital units that gives rise to the units we observe. Particularly, R-MITS does not account for heterogeneity of change points across units for situations where the data warrants such treatment. We will develop an interrupted time series mixed effects model as an alternative to R-MITS, able to detect unit specific change-points and borrow information across units while allowing for change point heterogeneity.

\bibliographystyle{unsrt}
\bibliography{ITS2B}

\section*{Acknowledgments}

This work was supported in part by the Eugene Cota-Robles Fellowship at the University of California, Irvine, the National Science Foundation Graduate Research Fellowship Program under Grant No. DGE-1321846, the National Science Foundation MMS 1461534 and DMS 1509023 grants, and by the National Institute on Aging of the National Institutes of Health under award numbers R01AG053555 and P50AG16573.
Any opinion, findings, and conclusions or recommendations expressed in this material are those of the authors(s) and do not necessarily reflect the views of the National Science Foundation. The content is solely the responsibility of the authors and does not necessarily represent the official views of the National Institutes of Health.

\appendix
\section{Appendices}

\subsection{Estimators of the Mean Function Parameters}

The generalized least squares (GLS) estimators for the mean function parameters of $\theta_1, \dots, \theta_J$ given $q \in Q$, obtained in step (6.) of Algorithm 1 of Section \ref{CPSection}, are
$$ 
\widehat{\theta}_j = \begin{bmatrix}
\widehat{ \beta}_{j0}  \\ 
\widehat{\beta}_{j1} \\
\widehat{\delta}_j \\ 
\widehat{\Delta}_j
\end{bmatrix} = \big[ \, {\bm X}(q)_j^{'} \, \,  \widehat{ \Sigma}_{j}^{-1} \, \,  {\bm X}(q)_j \, \big]^{-1}  \big[ \, {\bm X}(q)_j^{'} \, \,  \widehat{ \Sigma}_{j}^{-1}  \,\, Y_{j} \, \big],
$$  

$$
\text{where     }  \,\,\,\,\,\,\,\,\,\,\, {\bm X( q ) }_j \, \,\equiv \begin{bmatrix}
1 & 2 &  0 & 0 \\
\vdots & \vdots & \vdots & \vdots \\
1 & q -1 & 0 & 0 \\
1 & q & 1 &q \\
\vdots & \vdots & \vdots & \vdots \\
1 & T & 1  & T 
\end{bmatrix},
$$
and  $\widehat{ \Sigma}_{j}$ given in the subsequent section. Hence, for unit $j$ the estimator of (a.) the intercept pre-change point is $\widehat{ \beta}_{j0}$; (b.) the slope pre-change point is $\widehat{ \beta}_{j1}$; (c.) the change in level (post-change point intercept anchored at $\widehat{\tau}$) is $\widehat{\delta}_j +  \widehat{\Delta}_j \, \widehat{\tau}$; and the change in slope is $\widehat{\Delta}_j.$ 
\vspace{-.4cm}

\subsection{Estimators of the AR(1) Processes Parameters}

In steps (4.) and (7.) of the iterative estimation process, provided in Section \ref{CPSection}, the residuals, $r_{jt} = y_{jt} - \widehat{\mu}_{jt}$, are modeled as AR(1) processes:
\begin{equation*}
r_{jt} = \left\{
\begin{array}{lrr}
\phi_{j1} \,\,  r_{j, t-1} + e_{jt, 1},  &  & \,\, \, \,  \, \, \,\, \, \,  \, \, \,  \, \, \, \, \, \, \, \, \,  \, \, \, 1 < t < q,\\
\phi_{j2}  \, \, r_{j, t-1} + e_{jt, 2},  &   &  q \leq t \leq  T, 
\end{array}
\right.
\end{equation*}
with $e_{jt, i} \stackrel{iid}{\sim} N \big( 0, \sigma_{jw, i}^2 \big)$ for $i \in \{1, 2\}.$ Recall, to ensure causality in the time series sense, $\phi_{j1}$ and $\phi_{j2}$ must lie in the interval $(-1, 1)$ for all $j.$
The variance-covariance matrix, ${\bm \Sigma}_j,$ is therefore equal to
\setlength{\arraycolsep}{3.5pt}
\medmuskip = 1mu
\small{
	$$ \left[\begin{array}{c@{}c}
	\dfrac{\sigma_{jw, 1}^2}{1 - \phi_{j1}^2}
	\begin{bmatrix}
	1 & \phi_{j1} & \dots & \phi_{j1}^{q -2} \\
	\phi_{j1} & 1 & \dots &  \phi_{j1}^{q -3} \\
	\vdots & \vdots & \ddots & \vdots \\
	\phi_{j1}^{q -2} & \phi_{j1}^{q -2} & \dots & 1  
	\end{bmatrix} & \bigzero \\[15pt]
	\bigzero & \dfrac{\sigma_{jw, 2}^2}{1 - \phi_{j2}^2}
	\begin{bmatrix}
	1 & \phi_{j2} & \dots & \phi_{j2}^{T - q} \\
	\phi_{j2} & 1 & \dots &  \phi_{j2}^{T - q -1} \\
	\vdots & \vdots & \ddots & \vdots \\[5pt]
	\phi_{j2}^{T - q}  &  \phi_{j2}^{ T - q-1} & \dots & 1  
	\end{bmatrix}
	\end{array}\right] 
	$$ }
\normalsize
\nopagebreak 
and completely determined by $\phi_{j1}, \, \phi_{j2}, \, \sigma_{jw, 1}$ and $\sigma_{jw, 2}.$ We therefore only provide the estimators of  $\phi_{j1}, \, \phi_{j2}, \, \sigma_{jw, 1}$ and $\sigma_{jw, 2},$ conditional on $q \in Q.$ 
 
Define
\begin{align*}
\overline{r}_{(1a)} &= \frac{1}{q-2} \sum\limits_{t=1}^{q-2} r_{t}, & \overline{r}_{(1b)} = \frac{2}{q-1} \sum\limits_{t=2}^{q-1} r_{t}, \\[5pt] 
\overline{r}_{(2a)} &= \frac{1}{T-q-1} \sum\limits_{t=q -1}^{T-1} r_{t}, & \overline{r}_{(2b)} = \frac{1}{T -q-1} \sum\limits_{t=q}^{T} r_{t}, \\[5pt]
\sigma_{r_1}^2 &=  \dfrac{\sum\limits_{t=2}^{q -1} (r_{t} - \overline{r}_{(1b)})^2 \, + \, \sum\limits_{t=2}^{q -1} (r_{t-1} - \overline{r}_{(1a)})^2 }{2}, &  \\[5pt]
\text{and} \,\,\,\,  \sigma_{r_2}^2 &=  \dfrac{\sum\limits_{t=q}^{T} (r_{t} - \overline{r}_{(2b)})^2 \, + \, \sum\limits_{t=q}^{T} (r_{ t-1} - \overline{r}_{(2a)})^2 }{2}. &
\end{align*}

{ \bf The estimators of  $\phi_{j1}, \, \phi_{j2}, \, \sigma_{jw, 1}, \, \sigma_{jw, 2}, \, \sigma_{j, 1}$, and $\sigma_{j, 2},$ conditional on $q,$ are }
\begin{itemize}
	\item $	\widehat{\phi}_{j1} = \dfrac{ \sum\limits_{t=2}^{q-1} (r_{t} - \overline{r}_{(1b)}) \, (r_{t-1} - \overline{r}_{(1a)})}{ \sigma_{r_1}^2} $ 
	
	\item $\widehat{\phi}_{j2} = \dfrac{ \sum\limits_{t=q}^{T} (r_{t} - \overline{r}_{(2b)}) \, (r_{t-1} - \overline{r}_{(2a)})}{ \sigma_{r_2}^2}$
		
	\item 	$\widehat{\sigma}_{jw, 1}^2 =  \tfrac{1}{q-2} \,\, \sum\limits_{t=2}^{q-1} \, \Big[ (r_{t} - \overline{r}_{(1b)}) - \widehat{\phi}_{j1} ( r_{t-1} - \overline{r}_{(1a)}) \Big]^2$
	
	\item$\widehat{\sigma}_{jw, 2}^2 =  \tfrac{1}{T- q +1} \,\, \sum\limits_{t=q}^{T} \, \Big[ (r_{t} - \overline{r}_{(2b)}) - \widehat{\phi}_{j2} ( r_{t-1} - \overline{r}_{(2a)}) \Big]^2$
    
     \item $\widehat{\sigma}_{j, 1} = \dfrac{\widehat{\sigma}_{jw, 1} }{\sqrt{ 1 - (\widehat{\phi}_{j1})^2 }}$ 
     
    \item $\widehat{\sigma}_{j, 2} = \dfrac{\widehat{\sigma}_{jw, 2} }{\sqrt{ 1 - (\widehat{\phi}_{j2})^2 }}.$ 
\end{itemize}

\subsection{Covariance Matrix of the Full Model Mean Function Parameters for the Supremum Wald Test}

The supremum Wald statistic of section \ref{Wald} depends on $ { \bf \widehat{V}}_{0}(\widehat{\vec{\beta}}^{0}),$ the block diagonal estimator of the variance covariance matrix of $\widehat{\vec{\beta}}^{0}.$ Each block of $ { \bf \widehat{V}}_{0}(\widehat{\vec{\beta}}^{0})$ corresponds to ${\bf \widehat{V}} (\widehat{\vec{\beta}}_j^{0} ),$ the estimated variance-covariance matrix of the mean function parameters for unit $j.$  
Note, $${\bf \widehat{V}} (\widehat{\vec{\beta}}_j^{0} ) =   \, \big( {\bf X}^{1 \, '} \,  (\widehat{ {\bf \Sigma}}_{ j}) ^{-1} \, {\bf X}^1 \big)^{-1},$$ with ${\bf X}^1$ as
the design matrix of the {\bf full} model (model of equation (\ref{NullModels})) and the variance-covariance matrix under the {\bf reduced} model (model of equation (\ref{FullModels})) as $\widehat{\bf{\Sigma}}_{ j}$. 
Since the aim of the supremum Wald test is to test the existence of a change point in the mean, we assume an autocorelation structure that remains constant over the entire duration of the observational period.
Thus, for unit $j$
$$
\widehat{\bf{\Sigma}}_{ j} = \frac{(\widehat{\sigma}_{jw})^2}{1 - (\widehat{\phi}_j)^2}
\begin{bmatrix}
1 & \widehat{\phi}_j & \dots & (\widehat{\phi}_j)^{T-2} \\
\widehat{\phi}_j & 1 & \dots &  (\widehat{\phi}_j)^{T-3} \\
\vdots & \vdots & \ddots & \vdots \\
(\widehat{\phi}_j)^{T-2}  &  (\widehat{\phi}_j)^{T-3} & \dots & 1  
\end{bmatrix},
$$
where $ \widehat{\phi}_j$ and $(\widehat{\sigma}_{jw})^2$ are estimated under the reduced model.

\end{document}